\documentclass[a4paper,11pt]{article}
\pdfoutput=1 

\usepackage{jcappub} 

\usepackage[T1]{fontenc} 

\usepackage{ae,aecompl}
\usepackage{graphicx}
\usepackage{amsmath} 
\usepackage{amssymb}
\usepackage{multicol}        
\usepackage{bm}		
\usepackage{pdflscape}	
\usepackage{verbatim} 
\usepackage{mathrsfs}
\usepackage{amsfonts}
\usepackage{color}
\usepackage{hyperref}
\usepackage{subeqnarray}
\usepackage{setspace}
\usepackage{widetable}

\graphicspath{ {../scripts/} {../../scripts/} {../../../PdeK/} {../../../Cls/} }

\newcommand{\bra}{\langle}
\newcommand{\ket}{\rangle}
\newcommand{\mH}{\mathcal{H}}
\newcommand{\nk}{\textbf{k}}

\newcommand{\dphi}{\delta \phi}
\newcommand{\mR}{\mathcal{R}}
\newcommand{\nq}{\textbf{q}}
\newcommand{\mP}{\mathcal{P}}
\newcommand{\x}{\textbf{x}}
\newcommand{\fin}{\textrm{end}}

\newcommand{\di}{\diamond}
\newcommand{\eidi}{\epsilon_{1 \diamond}}
\newcommand{\eiidi}{\epsilon_{2 \diamond}}

\newcommand{\nn}{\nonumber \\}
\newcommand{\RI}{\text{R,I}}
\newcommand{\nrun}{n_{\textrm{run}}}

\title{Observational constraints on the second-order primordial power spectrum: Exploring a Continuous Spontaneous Localization inspired inflationary model}

\author[a,b]{Mar\'{\i}a P\'{\i}a Piccirilli}

\author[a,b]{Gabriel Le\'{o}n,}

\affiliation[a]{Grupo de Cosmolog\'{\i}a, Facultad de Ciencias Astron\'{o}micas y Geof\'{\i}sicas, Universidad Nacional de La Plata, Paseo del Bosque S/N 1900 La Plata, Argentina}
\affiliation[b]{CONICET, Godoy Cruz 2290, 1425 Ciudad Aut\'onoma de Buenos Aires, Argentina}

\emailAdd{mpp@fcaglp.unlp.edu.ar }
\emailAdd{gleon@fcaglp.unlp.edu.ar}

\abstract{   Inflation, a period of exponential expansion in the early Universe, is considered an important part of the standard $\Lambda$CDM cosmological  model, and plays a crucial role in explaining a wide range of current observations. The standard inflationary model predicts a primordial spectrum of fluctuations that is nearly scale-independent, fitting remarkably well  the latest  observational data.  
	Nevertheless, there is an ongoing discussion surrounding the transition from an initial homogeneous and isotropic quantum state, characterizing the matter fields during inflation, to a classical inhomogeneous/anisotropic one, which gives rise to large-scale structure in the Universe.
	To tackle this issue, in the present work we explore an inflationary scenario where quantum ``collapse'' (or reduction) occurs naturally during the  evolution of the system; this model is inspired in the so called Continuous Spontaneous Localization (CSL) model. Our present work builds upon previous results by considering the primordial power spectrum up to the second order in the Hubble Flow Functions, where we perform an estimation of the model free parameters. By validating the predictions of the model against observational data, we investigate whether this second-order calculation can explain the slight departure from the power law observed in the scalar spectral running index. We hope this research contributes to the understanding of the quantum-to-classical transition and its implications for cosmology. }

\begin{document}
\maketitle
\flushbottom

\section{Introduction}

{The cosmological model, according to which our Universe has been in a state of perpetual expansion, is supported by observational evidence, and modeled within the theoretical framework of General Relativity \citep{Friedmann, Einstein, deSitter}.  This evidence was first  historically  established by Hubble \citep{Hubble} and Lema\^itre\footnote{If the Universe was expanding, then it should have been smaller in the past. This led Lema\^itre to propose the {\em primeval atom} hypothesis, which laid the foundations for what we now call the {\em Big Bang} singularity.} \citep{Lemaitre}, the latter employed measurements of velocity spectra of extra-galactic nebulae \citep{Slipher}, which can be used to estimate distances using the relationship involving cepheids \citep{Henrietta}. Currently, the data supporting the expansion of the universe includes (among others): distance ladder measurements of local type Ia supernovae (SNe) \citep{Feeney2017}, measurements of the Cosmic Microwave Background (CMB) \citep{Planck18c}, and baryon acoustic oscillation (BAO) measurements of angular-diameter distances \citep{eBOSS}.  On the other hand, inflation, a period characterized by an exponential expansion in the early Universe (prior to the decoupling epoch) \citep{Guth81, Hawking82, Mukhanov}, has become an integral part of modern cosmology, in great part due to its empirically successful account of the primordial inhomogeneities that represent the origin of the cosmological structure. All these features together give rise to what is known as the concordance $\Lambda$CDM cosmological model. In fact, observational data is becoming increasingly abundant and unprecedentedly accurate, consistently reaffirming the accepted status  of the {\em Big Bang} model + inflation, within the combined framework of General Relativity and Quantum Theory.\footnote{Here, we imply that at the fundamental level, the standard cosmological model treats spacetime as dictated by the equations of General Relativity, and the matter fields as described by Quantum Field Theory. } } 

{ Observations from the CMB \citep{penziaswilson, cobe, wmap9cls, Planck18a} provide compelling evidence for the highly isotropic cosmic radiation that we observe today; this radiation represents a relic from the early stages of the Universe. Additionally, 
	recent CMB data \citep{Planck18a, Planck18b, Planck18c} are consistent with the predictions of standard inflation, which entail a primordial spectrum that is highly Gaussian, adiabatic, and nearly scale-invariant. However, some aspects of the conventional inflationary paradigm, particularly concerning a clear explanation for the generation of primordial homogeneities, remain unsettled. Specifically, an open problem within standard inflation, initially mentioned in \citep{PSS06} and subsequently analyzed in other works like \citep{Shortcomings, Susana13, Okon16, Elias2021}, is as follows: } The inflaton, i.e. the scalar field responsible for driving the exponential expansion, initially exists in a perfectly homogeneous and isotropic quantum state.\footnote{  {In the standard inflationary model, the  initial state of the inflaton is commonly characterized as a highly excited state for the zero mode of the field, and the so called Bunch-Davies vacuum for the rest of the field modes.} } However, none of the quantum mechanical equations used to evolve the state of the inflaton have the potential to break these symmetries (homogeneity and isotropy). This raises the fundamental question: How do primordial inhomogeneities, which eventually give rise to large-scale structure, emerge from the quantum state of the inflaton that characterizes the matter fields in the primordial Universe?  A further related question that ensues is: How is it possible for the Universe to contain classical galaxies considering its quantum origin? Some may argue that decoherence has the potential to address the latter question \citep{kiefer2,polarski}. However, we contend that it is insufficient to answer the former question \citep{Shortcomings, Okon16}.

This debate is not new, and a large amount of literature has been dedicated to these concerns \citep{jmartin, jmartinPRL, pedrocsl, Shortcomings, Das2013, kiefer, polarski, pintoneto, valentini, goldstein, Stephon16, Picci19, Leon2020, Leon2021,  ashtekar2020}. The reason why it remains an open discussion lies in the fact that the quantum treatment of the Universe inherits unresolved matters from quantum mechanics, specifically the measurement problem \citep{Bell81,Maudlin95}. Concretely, when the entire Universe is the subject of study, there is no transparent and global way to make a distinction between the ``observer'', or measurement apparatus, and the ``observed object'' or physical system. In fact, when applying Quantum Theory to the early Universe, what we are trying to explain are the primordial conditions that, eventually, led to our existence. Evidently, it is senseless to allow for considerations regarding what we, humans, can or cannot in fact measure, to play a role in such an explanation \citep{Elias2021}. It also challenges the notion of constructing a ``super-observer'' outside the Universe, which is clearly not a viable solution within any reasonable physical theory.

The idea that the quantum wave function spontaneously ``collapses'' during its dynamical evolution has been put forward as a possible solution to the measurement problem \citep{Ghirardi86, Pearle89, Diosi84, Penrose96}. This proposal belongs to a broader class of models known as objective reduction models or simply ``collapse models'' \citep{Bassi1}. One of the most extensively studied and tested models in this category is the Continuous Spontaneous Localization (CSL) model \citep{Bassi2, Pearle1995, PearleMisc, Pearle06A}.

The CSL model introduces a modification to the {Schr\"o\-din\-ger} equation by incorporating a stochastic term, and provide also a Probability Rule, which gives the probability of a particular realization of the stochastic term. All these elements leads to the collapse of the state vector into a new state, which in principle does not share the same symmetries as the initial state. The CSL model has also been subject to experimental testing as its predictions differ from those of standard quantum mechanics for certain specific systems \citep{Bassiexp21, Bassi2022}.

Furthermore, the CSL model can be extended to the cosmological context, and in particular, has been successfully applied to the inflationary period, where it has demonstrated its ability to fit actual observational data \citep{Das2013, pedrocsl, BassiCMB, Leon16, Maj17, Picci19, Beneficios,  jmartin, jmartinPRL, Bassi2021, Leon2020,  Leon2021}. This demonstrates that the aforementioned problem cannot be dismissed as merely philosophical in nature. Instead, the attempts to address this issue have tangible effects on the predictions of the model, which can be tested against recent observational data.

On the other hand, several previous studies have examined the theoretical predictions of the traditional inflationary paradigm when the primordial power spectrum was calculated up to the first order in the Hubble Flow Functions (HFF), e.g. \citep{jmartinpotentials, jmartin2019, Planck18b}. However, latest data reveals certain features that cannot be overlooked when constructing an inflationary cosmological model \citep{Adshead2010, Bahr2021}. Specifically, the data seem to indicate that a scale dependence of the scalar spectral index $n_s$ is still allowed; this scale dependence of $n_s$ is known as the running of the spectral index  $n_{\rm run}$.  In other words, the data admits a slight deviation from the power law behavior in the primordial power spectrum. As observed in \citep{Planck15_inf, Planck18b}, the parameter $n_{\rm run}$  is consistent with a value of zero, but it is not centered precisely at zero. There exists a range of values that are equally probable such that $ |n_{\rm run}| \lesssim 10^{-3}$ \citep{Planck18b}, and the possibility of a non-zero value must be taken into account in future experiments \citep{Bahr2022}.

As a matter of fact, future observational constraints on $n_{\rm run}$ can be considered as a probe of the traditional single-field slow-roll inflationary model \citep{Adshead2010, Bahr2022}. As is well known, to the lowest order of the HFF, the predictions derived from slow roll inflation are as follows: $n_s -1 = -2 \eidi - \eiidi$, $r= 16 \eidi$ (referred to as the tensor-to-scalar ratio), and $n_{\rm run} = -2 \eidi \eiidi - \eiidi \epsilon_{3 \di}$, where $\epsilon_{j \di}$ denotes the evaluation of the HFF at the pivot scale. Consequently, in slow-roll inflation all of the previous parameters are related as $n_{\rm run} = (n_s-1 + r/8)(r/8 + \epsilon_{3 \di})$.

Recent data have narrowed the value of the scalar spectral index $|n_s - 1 | \simeq 10^{-2}$ \citep{Planck18b}. However, a plausible future scenario is that primordial gravitational waves remain undetected, resulting in tighter bounds on $r$, which means lower values allowed by the data for $\epsilon_{1 \diamond}$. In that case, $n_s$ and $n_{\rm run}$, would be the main parameters used to discriminate among possible inflationary models. Furthermore, if future experiments are consistent with a high value for the running parameter, i.e. $|n_{\rm run}| \simeq 10^{-3}$, then it would imply $|\epsilon_{3 \diamond}| > |\epsilon_{2 \diamond}| > |\epsilon_{1 \diamond}|$. Thereby contradicting the hierarchy of the HFF \citep{liddleSR}, and suggesting a potential inconsistency with standard slow-roll inflation \citep{runninglewis}.

In Ref. \citep{running2020}, we calculated the primordial power spectrum at the next leading order in the HFF that results from applying a CSL (inspired) inflationary model within the semiclassical gravity framework, which we dubbed CSLIM for simplicity. The CSLIM predicts a strong suppression of primordial gravity waves generated by second-order scalar perturbations \citep{Lucila15,Maj17}. The estimated tensor-to-scalar ratio in the CSLIM is $r \simeq 10^{-7} \epsilon_{1 \diamond}^2$, which can be seen is decoupled from $n_s-1$ and $n_{\rm run}$; this is a major difference with respect to the standard prediction. Also, the spectrum in CSLIM exhibits an additional $k$ dependence through the function $C(k)$, potentially acting as an additional ``running effect'' independent of $n_{\rm run}$. Therefore, such a feature may prove useful for the resolution of inconsistencies within the slow-roll inflationary model, without violating the hierarchy of the HFF if future experiments detect a value of  $|n_{\rm run}| \simeq 10^{-3}$, which we repeat is not too unrealistic. Additionally, this showcases that the CSLIM holds important observational consequences, hence the issue mentioned previously cannot be dismissed as purely philosophical.

In this article we perform a statistical analysis to compare the predictions obtained using the CSLIM with recent CMB data. In particular, we focus on the previously calculated second-order primordial spectrum \citep{running2020} and aim to investigate whether small deviations from the conventional power law in the power spectrum are necessary or if they naturally arise as a consequence of the CSLIM. 

The paper is organized as follows: In Sec. \ref{Sec2}, we present a summary of the CSL inspired model and its implementation in the inflationary epoch; here we also state the main assumptions adopted to obtain the primordial power spectrum at second order in the HFF. In Sec. \ref{Sec3}, we use the observational data from the CMB to constrain the cosmological parameters and estimate the free parameter of the CSL inflationary model. We also perform a comparison between the standard $\Lambda$CDM model and our proposed model. In Sec. \ref{Sec4}, we conduct a secondary analysis using observational data to test our main hypothesis, which is to determine whether the power spectrum obtained from the CSL inflationary model can mimic the effects of the running of the spectral index. This investigation aims to address the potential future tension in the standard slow roll inflationary scenario. Finally, in Sec. \ref{Sec5}, we present our conclusions. Regarding conventions and notation, we use the metric signature $(- + + +)$. We choose units such that $c=\hbar=1$, but maintain the gravitational coupling constant $G$. An overdot is used to denote derivatives w.r.t. cosmic time $t$, whereas a prime is used for derivatives w.r.t conformal time $\eta$.

%

%




\section{The CSL inflationary model}\label{Sec2}
The purpose of this section is to provide a brief overview of the calculations developed in \citep{running2020}. Therefore, most of the details are omitted and no original work is presented in this section.

We begin by establishing the foundation of our proposed model within the semiclassical gravity framework, where gravity is treated classically while matter fields are analyzed from a quantum field theory perspective. This approach can be regarded as an effective theory that describes quantum matter fields in a classically curved spacetime. Consequently, the semiclassical Einstein's equations are given by
\begin{equation}\label{scEE}
	G_{ab}  = 8 \pi G \bra \hat T_{ab} \ket,
\end{equation}
where $G_{ab}$ is Einstein's tensor and $\bra \hat T_{ab} \ket$ represents the quantum expectation value of the energy-momentum tensor of the matter fields.

\subsection{Slow roll model}

The energy-momentum tensor on the right-hand side of \eqref{scEE} corresponds to the inflaton, which in its simplest formulation is a single scalar field with canonical kinetic energy in some initial state (the Bunch-Davies vacuum state). The standard action for the inflaton involves a single scalar field minimally coupled to gravity:
\begin{equation}\label{action0}
	S[\phi,g_{ab}] = \int d^4x \sqrt{-g} \bigg[ \frac{1}{16 \pi G} R[g] 
	- \frac{1}{2}\nabla_a \phi \nabla_b \phi g^{ab} - V[\phi] \bigg].
\end{equation}
Here, $V[\phi]$ represents an appropriate potential that is determined empirically. Hence, the matter sector consists  of a scalar field $\phi(\x,t)$, which will be treated perturbatively by decomposing it into a homogeneous part and a small perturbation: $\phi_0(t) + \delta\phi(\x,t)$. The metric $g_{ab}$ is also split into a homogeneous and isotropic background $g_{ab}^{(0)}$, which corresponds to a flat Friedman-Lema\^{\i}tre-Robertson-Walker (FLRW) spacetime with a scale factor $a(t)$, and small inhomogeneous perturbations $\delta g_{ab}$.

Slow-roll inflationary models rely on a set of parameters that determine the characteristics of the inflationary period. These parameters can be defined in terms of the Hubble parameter $H \equiv \dot{a}/a$ (where the dot denotes the derivative with respect to cosmic time $t$) to yield what we refer to as the \textit{Hubble flow functions} (HFF) \citep{terreroHFF,terreroHFF2}:
\begin{equation}\label{defepsilonn}
	\epsilon_{n+1} \equiv \frac{d \ln \epsilon_n}{d N}, \qquad \epsilon_0 \equiv \frac{H_{\text{ini}}}{H},
\end{equation}
where $N \equiv \ln (a/a_{\text{ini}}) $ is the number of e-folds from the beginning of inflation.

The inflationary period occurs only if $\epsilon_1 < 1$, and the slow-roll approximation assumes that $|\epsilon_n| \ll 1$, meaning that all HFF parameters remain small throughout the duration of inflation.

The Friedmann equations for the background matter fields can be expressed in terms of the first two HFFs:
\begin{equation}\label{friedamnnSR}
	H^2 = \frac{V}{ M_P^2 (3-\epsilon_1)},
\end{equation}
\begin{equation}\label{KGSR}
	3H \dot \phi \left(1-\frac{\epsilon_1}{3} + \frac{\epsilon_2}{6} \right) = - \partial_\phi V,
\end{equation}
where $V = V(\phi_0)$  and $ M_P^2 \equiv 1/(8\pi G)$ is the reduced Planck's mass. It is important to note that these equations are exact.

Next, we turn our attention to the perturbations in the theory. Regarding the metric perturbations, we opt to work in the Newtonian (or longitudinal) gauge. Additionally, we introduce a change of the time coordinate by setting $dt = a(\eta) d\eta$, where $\eta$ represents the conformal time. Consequently, the line element associated with the metric at first order in the perturbations can be expressed as:
\begin{equation}
	ds^2 = a^2(\eta) \left[ - (1+2\Phi) d\eta^2 + (1-2 \Psi) \delta_{ij} dx^i dx^j \right].
\end{equation}
In this equation $i,j = 1,2,3$;  $\Phi$ and $\Psi$ represent the scalar perturbations of the metric. In fact, using  Einstein's Equations (EE) at linear order and assuming no anisotropic stress, one finds that $\Phi = \Psi$. We will refer to $\Psi$ as the  {\em Newtonian potential}, which represents the so called curvature perturbation\footnote{Here, \textit{curvature perturbation} means the intrinsic spatial curvature on hypersurfaces of constant conformal time, for a spatially flat universe.} in the Newtonian gauge. 


In Ref. \citep{running2020}, we have shown that, using the semiclassical EE at linear order in perturbations and definitions \eqref{defepsilonn}, the main equation relating the metric and the inflaton perturbations are given, in Fourier's space, by
\begin{equation}\label{masterpsi}
	\Psi_{\nk} \simeq \frac{  1  }{ M_P }  \sqrt{ \frac{\epsilon_1}{2 } } \frac{\bra \hat \dphi_{\nk} \ket }{ (1+\epsilon_2) } ,
\end{equation}
which is valid only up to second order in $\epsilon_n$. The previous equation clearly demonstrates the distinctive contribution of our proposal. When the initial state of the quantum field is taken to be the traditional Bunch-Davies vacuum, which is perfectly isotropic and homogeneous, the right-hand side of \eqref{masterpsi} becomes exactly zero. As a result, no perturbations of any specific scale are present in the Universe. However, after the self-induced collapse of the wavefunction has occurred (provided by the CSL mechanism), the final state no longer possesses the same symmetries as the initial Bunch-Davies vacuum state. This implies that the right-hand side of \eqref{masterpsi} is no longer zero, and thus $\Psi_{\nk} \neq 0$, giving rise to the primordial perturbation. For further discussions on this issue, Refs. \citep{alberto,erandy,benito,benito2019} provide more extensive conceptual and technical insights.

{Lastly, it is important to recall that in our framework, the metric is considered a classical variable, effectively describing the deeper fundamental degrees of freedom of an underlying quantum gravity theory. Meanwhile, matter fields, specifically the inflaton perturbations, receive a standard quantum field treatment in curved spacetime. These two aspects are related through the semiclassical Einstein's equations. In contrast, the standard approach in inflationary models involves the quantization of both $\Psi$ and $\dphi$, which are then linearly combined to form what is called the Mukhanov-Sasaki variable $v$ \citep{Mukhanov81,Mukhanov}. Subsequently, the quantum field theory of $v$ is analyzed.}

{The Newtonian gauge proves to be highly valuable within the semiclassical gravity framework for conducting calculations. As illustrated in Eq. \eqref{masterpsi}, this gauge explicitly unveils the distinct fundamental basis of spacetime and matter fields in our approach. Moreover, once we establish this key equation, which relates the metric and matter perturbations through the semiclassical gravity formalism (while utilizing the CSL mechanism), we can seamlessly transition to any chosen gauge or use it to construct gauge-invariant quantities. In the upcoming section, we will switch to the comoving gauge because it enables us to compare our predictions with the standard ones, typically expressed in the comoving gauge.}

\subsection{Primordial Power Spectrum}

Once we have obtained the expression for the metric perturbation \eqref{masterpsi}, our next objective is to derive the scalar power spectrum in order to establish a connection between our theoretical proposal and its observational predictions.

We begin by emphasizing that the Newtonian potential $\Psi$ is a conserved quantity for super-Hubble scales. Although its amplitude may change during different cosmological epochs. For instance, during the transition from the radiation to the matter-dominated era, this variation is proven to be negligible, but significant from inflation to the radiation epoch (see \citep{Daniel10,running2020}).

Furthermore, there exists another conserved quantity for adiabatic perturbations on super-Hubble scales that remains unaffected by the cosmological epoch:
\begin{equation}\label{relacionRyPsiexacta}
	\mR \equiv \Psi  + \left(  \frac{2 \rho}{3}   \right) \left( \frac{\mH^{-1}  \Psi' + \Psi}{\rho + P}\right),
\end{equation}
where $\rho$ and $P$ are the energy and pressure densities associated to the type of matter driving the expansion of the Universe; the prime denotes derivative with respect to $\eta$. This quantity represents the curvature perturbation in the comoving gauge and is associated with the primordial power spectrum in the standard approach. Specifically, in Fourier space, the primordial power spectrum of the comoving curvature perturbation is:
\begin{equation}\label{PSdef}
	\overline{\mR_{\nk}\mR^*_{\nq}} \equiv \frac{2 \pi^2}{k^3} \mP_{s} (k) \delta(\nk-\nq),
\end{equation} 
where $\mP_{s} (k)$ is the dimensionless power spectrum and the bar denotes an ensemble average over possible realizations of the stochastic field $\mR_{\nk}$.

By considering the exact relation \eqref{relacionRyPsiexacta} between $\Psi$ and $\mR$, we can calculate the curvature perturbation in the Newtonian gauge within the collapse picture, and subsequently switch to the comoving gauge for comparison with the standard approach.

The scalar power spectrum associated with $\mR_{\nk}$, derived from the expressions \eqref{masterpsi} and \eqref{relacionRyPsiexacta}, is given by:
\begin{equation}\label{PScolapso}
	\overline{\mR_\nk \mR_\nq^*} =  \frac{1}{2 M_P^2 \epsilon_1 }  \frac{(1+ \epsilon_1 + \epsilon_2)^2}{(1+\epsilon_2)^2}  \overline{\bra \hat \dphi_\nk \ket \bra \hat \dphi_\nq \ket^*},
\end{equation}
where in the context of the CSLIM, each realization is associated with a particular realization of the stochastic process that characterizes the collapse process.

Comparing equations \eqref{PSdef} and \eqref{PScolapso}, the scalar power spectrum can be identified as
\begin{equation}\label{masterPS}
	\mP_{s} (k) \delta(\nk-\nq) = \frac{k^3}{4 \pi^2 M_P^2 \epsilon_1 }  \frac{(1+ \epsilon_1 + \epsilon_2)^2}{(1+\epsilon_2)^2}  \overline{\bra \hat \dphi_\nk \ket \bra \hat \dphi_\nq \ket^*}.
\end{equation}

The next step is to employ the CSLIM  to obtain \\$\overline{\bra \hat{\dphi}_\nk \ket \bra \hat{\dphi}_\nq \ket^*}$ in the super-Hubble regime $k \eta \to 0$.

\subsection{Quantization}


We then proceed to quantize the perturbation $\dphi(\x,\eta)$ in a curved quasi-de Sitter spacetime background. Detailed calculations can be found in \citep{running2020}, but for the sake of completeness, we will summarize the major steps in this subsection.

We begin by expanding the action \eqref{action0} up to second order in the perturbations $\dphi$ and $\Psi$. Since we are working within a semiclassical framework, only the matter degrees of freedom will be quantized. By rescaling the field variable to $y=a\dphi$ and considering that $S = \int d^4x \delta^{(2)} \mathcal{L}$, we obtain:
\begin{eqnarray}\label{action2}
	\delta^{(2)} \mathcal{L} &=& \frac{1}{2} \bigg[  y'^2   - (\nabla y)^2  - y^2 a^2  V_{, \phi \phi} +  \frac{a''}{a} y^2\bigg] \nonumber \\
	&+& a [4\phi_0' \Psi' y - 2 a^2 V_{, \phi}  \Psi y  ],
\end{eqnarray}
where $V_{, \phi}$ indicates partial derivative with respect of the field $\phi$.

According to our approach,  $\Psi = \Psi' =0$ in the vacuum state.  However, the CSL mechanism implies a continuous process evolving the initial vacuum state $| 0 \ket$ into a different state, e.g. $| \Xi \ket$. Consequently, the metric perturbations will also change from zero to a non-vanishing value in a continuous manner. Therefore, the  terms containing $\Psi$ and $\Psi'$ in \eqref{action2} can be considered as a backreaction effect of the CSL mechanism, which is of second order in HFF. In Fourier's space, the action is $ S = \int d\eta  {L}^{(2)}$, with\\ ${L}^{(2)} \equiv \int_{\mathbb{R}^{3+}} d^3 k  \mathcal{L}^{(2)}$ can be written as: 
\begin{eqnarray}\label{lagrangianok}
	\mathcal{L}^{(2)} &\equiv&   y_\nk' y_\nk^{*'} - (  k^2 - \frac{a''}{a} + a^2   V_{, \phi \phi}  ) y_\nk y_\nk^*  \nn
	& +& 4 a \phi_0' ( \Psi'_\nk y_\nk^* +  \Psi^{*'}_\nk y_\nk)\nn
	&-&2 a^3    V_{,  \phi} ( \Psi_\nk y_\nk^* + \Psi_\nk^* y_\nk). \nn
\end{eqnarray}
From the Lagrangian above, one can obtain the equation of motion for the field $y_\nk$,
\begin{equation}
	y_\nk'' + \left( k^2 - \frac{a''}{a}  +  a^2 V_{, \phi \phi}        \right)y_\nk   - 4 a \phi_0'  \Psi'_\nk   + 2 a^3    V_{,  \phi}  \Psi_\nk = 0.
\end{equation}

The CSL model is based on a non-unitary modification to the Schr\"odinger equation.  In this manner, it is convenient to proceed with the quantization in the Schr\"odinger picture, where one of the relevant physical objects is the Hamiltonian. 

The Hamiltonian associated to Lagrangian ${L}^{(2)}$, can be found as $H^{(2)} =  \int_{\mathbb{R}^{3+}} d^3k \: (y^{*'}_\nk p_\nk + y^{'}_\nk p^*_\nk   ) -  {L}^{(2)}$, where $y_{\nk}$ is $p_{\nk} \equiv
\partial \mathcal{L}^{(2)}/ \partial y_{\nk}^{\star '}$, that is $p_{\nk} =  y_{\nk}'$. Therefore, ${H}^{(2)} = \int_{\mathbb{R}^{3+}} d^3 k \mathcal{H}^{(2)}$, with
\begin{eqnarray}\label{Hamiltclas}
	\mH^{(2)} &\equiv&    p^*_\nk p_\nk  + y^*_\nk y_\nk \left( k^2 - \frac{a''}{a}  + a^2  V_{, \phi \phi}    \right)  \nn
	& -& 4 a \phi_0' ( \Psi'_\nk y_\nk^* +  \Psi^{*'}_\nk y_\nk) \nn
	&+& 2 a^3    V_{,  \phi} ( \Psi_\nk y_\nk^* + \Psi_\nk^* y_\nk). \nn  
\end{eqnarray}

Quantization is achieved by promoting $y_\nk$ and $p_\nk$ to quantum operators,  $\hat y_\nk$ and $\hat p_\nk$, and by requiring the canonical commutation relations,
\begin{equation}
	[\hat y_\nk  , \hat p_\nq ] = i \delta(\nk - \nq).
\end{equation}
Furthermore, we choose to work with  Hermitian operators; hence, we introduce the following definitions:
\begin{equation}
	\hat y_\nk \equiv \frac{1}{\sqrt{2}} (\hat y_\nk^\text{R} + i \hat y_\nk^\text{I}  ), \qquad \hat p_\nk \equiv \frac{1}{\sqrt{2}} (\hat p_\nk^\text{R} + i \hat p_\nk^\text{I}  ).
\end{equation}

In the Schr\"odinger picture, the quantum state of the system is described by a wave functional, $\Phi[y(\x,\eta)]$, which in Fourier space can also be factorized into mode components as
\begin{equation}
	\Phi[y(\x,\eta)] = \prod_{\nk}  \Phi_\nk ( y_\nk^\text{R} , y_\nk^\text{I} ) =  \prod_{\nk}  \Phi_\nk^\text{R} ( y_\nk^\text{R} )  \Phi_\nk^\text{I} ( y_\nk^\text{I} ).
\end{equation}


Thus, the wave functional $\Phi_\nk$ obeys the Schr\"odinger equation:
\begin{equation}\label{scheq}
	i \frac{\partial \Phi_\nk^\RI }{ \partial \eta} = \hat H^\RI_\nk  \Phi_\nk^\RI,
\end{equation}
where the Hamiltonian densities $\hat H^\RI_\nk $, are related to the Hamiltonian as $\hat{H}^{(2)} = \int_{\mathbb{R}^{3+}} d^3 k (\hat H^\text{R}_\nk  + \hat H^\text{I}_\nk ) $, with the following definitions
\begin{eqnarray}\label{HamiltRI}
	\hat H^{R,I}_\mathbf{k } &=& \frac{(\hat p_\mathbf{k }^{R,I} )^2  }{2 }+ \frac{(\hat y_\mathbf{k }^{R,I} )^2  }{2 }  \left( k^2 - \frac{a''}{a}  + a^2  V_{, \phi \phi}    \right)   \nn
	& -& 4 a \phi_0'  \Psi_\nk^{'\RI} \hat y_\nk^\RI  + 2 a^3    V_{,  \phi} \Psi_\nk^\RI \hat y_\nk^\RI. 
\end{eqnarray}

The standard assumption is that, at an early conformal time $\tau \to -\infty$, the modes
are in their adiabatic ground state, which is a Gaussian centered at zero with certain
spread. This ground state is commonly referred to as the Bunch-Davies (BD) vacuum. Therefore, the wave functional 
\begin{equation}\label{psionday}
	\Phi^{R,I}(\eta,y_{\nk}^{R,I}) = \exp[- A_{k}(\eta)(y_{\nk}^{R,I})^2 +
	B_{k}(\eta)y_{\nk}^{R,I} +  C_{k}(\eta)]
\end{equation}
evolves according to Schr\"odinger equation \eqref{scheq}, with initial 
conditions given by
\begin{equation}
	A_k (\tau ) = \frac{k}{2}, \qquad B_k (\tau ) = C_k(\tau )= 0,
\end{equation}
which characterizes the BD vacuum.

\subsection{CSL in the inflationary context}

The main physical idea  of the CSL model is that an objective reduction of the wave function occurs all the time for all kind of particles. The reduction or \textit{collapse} is spontaneous and random, taking place whether the particles are isolated or interacting and whether the particles constitute a macroscopic, mesoscopic or microscopic system \citep{Bassi1, Bassi2, Pearle06A}.

In order to apply the CSL model to the inflationary setting, we will consider a version of the CSL model adapted to inflation. In particular, we assume that  the objective reduction mechanism acts on each mode of the field independently.  The model is thus characterized by two equations:

\begin{eqnarray}\label{CSLevolution}
	|\Phi_{\nk}^{\textrm{R,I}}, \eta \ket &=& \hat T \exp \bigg\{ - \int_{\tau}^{\eta} 
	d\eta'   \bigg[ i \hat{H}_{\nk}^{\textrm{R,I}} \nn
	&+& \frac{1}{4 \lambda_k} (\mathcal{W}(\nk,\eta') - 2 \lambda_k
	\hat{y}_{\nk}^{\textrm{R,I}})^2 \bigg] \bigg\} |\Phi_{\nk}^{\textrm{R,I}}, \tau \ket, \nn
\end{eqnarray} 
and
\begin{equation}\label{cslprobabF}
	P(\mathcal{W}_{\nk}^{\RI}) d\mathcal{W}_{\nk}^{\RI} =   \bra \Phi_{\nk}^{\RI} , \eta | \Phi_{\nk}^{\RI}, \eta \ket \prod_{\eta'=\tau}^{\eta-d\eta} \frac{ d \mathcal{W}_{\nk}(\eta')^{\RI}}{\sqrt{2 \pi \lambda_k/d\eta}},
\end{equation}
where $\hat T$ is the time-ordering operator, and recall that $\tau$ denotes the conformal time at the beginning of inflation. The former equation is the evolution equation for the wave function, while the latter is the probability law.  The stochastic field $\mathcal{W}_\nk= \mathcal{W}_{\nk}^{\text{R}} + i \mathcal{W}_{\nk}^{\text{I}}   $ depends on $\nk$ and the conformal time.  In other words,  we have introduced a stochastic function for each independent degree of freedom, i.e. we are applying the CSL collapse dynamics to each mode of the field. Therefore, the stochastic field $\mathcal{W}_\nk (\eta)$ corresponds to a Fourier transform on a stochastic spacetime field $\mathcal{W}(\x,\eta)$ with probability given by \eqref{cslprobabF}.

From equation \eqref{CSLevolution}, it is clear that the field variable $\hat{y}_\nk^\RI$ has been chosen as the collapse generating operator, which means that the CSL process will drive the initial state vector towards an eigenstate of $\hat{y}_\nk^\RI$. This choice for the collapse operator is motivated by the fact that in Eq. \eqref{masterpsi}, the expectation value $\bra \hat{\delta \phi}_\nk \ket = \bra \hat{y}_\nk \ket/a$ acts as the source for the metric perturbation.

%
%

\subsection{Parameterization of $\mathbf{\lambda}_k$}

Equation  \eqref{CSLevolution} is continuous in Fourier space, and given the assumption 
that the reduction mechanism acts on each mode independently, the main CSL
parameter $\lambda_k$ will now depend on the mode $k$. The parameter $\lambda_k$ can also be considered as  a phenomenological generalization of the universal CSL parameter $\lambda_0$; this is, $\lambda_0^{-1}$ provides us with a localization time scale for the wave function associated to each mode of the field. From the point of view of pure dimensional analysis, $\lambda_k$ must have dimensions of [Length]$^{-2}$. Therefore, the simplest functional dependence is to choose
\begin{equation}\label{parametrizacion0}
	\lambda_k = \lambda k,
\end{equation}
{where $\lambda$ can be related in principle to the universal CSL rate parameter, the latter denoted as $\lambda_0$ with units of [Time]$^{-1}$. Other functional forms of $\lambda_k$ could be considered, for instance: $\lambda_k = \lambda^3/k$, affecting directly the shape of the primordial power spectrum \citep{pedrocsl}. However, as we will see in the next section, it is notable that the functional form proposed in \eqref{parametrizacion0} is not only the simplest one, but also compatible with the observational data.} 

{Another important aspect worth mentioning is that extrapolating laboratory estimations to the primordial Universe involves some caveats \citep{Leon2020}. In particular, we can identify two distinct regimes: the laboratory regime, where the CSL parameter $\lambda_0$ has been empirically constrained, and the inflationary era.  Consequently, there are ample reasons to question any straightforward connection between the value of $\lambda$, relevant for the inflationary regime (as it appears in Eq. \eqref{parametrizacion0}), and the parameter value  characterizing the theory in the laboratory. In fact, very different choices of the value of $\lambda$ could result in distinct predictions for the amplitude of the primordial spectrum\footnote{However, for a detailed analysis about this point, we invite the reader to check Sec. IV A of Ref. \citep{running2020}} \cite{pedrocsl}. }

{At this point, since we do not yet have a complete CSL theory, meaning a CSL-type equation that can be applied across all physical scales, we are left with making educated guesses. In Ref. \citep{Leon2020}, one of us analyzed the proposal that it is reasonable to consider the incorporation of aspects tied to spacetime curvature into the collapse parameter for cosmological systems, where the curvature is significant. Therefore, a potential extrapolation of the collapse parameter $\lambda$, to more general regimes involves assuming an explicit dependence of $\lambda$ on the spacetime curvature characterized by the Ricci scalar $R$ (although other choices are also possible, e.g. $R_{ab} R^{ab}$ or functions of the Weyl's scalar). One possible functional form\footnote{In fact the functional form \eqref{lambdaR} has been considered in Refs. \citep{Modak14, Modak15} where a possible resolution to the black hole information loss paradox, based on the CSL mechanism, was analyzed. } for $\lambda(R)$ is as follows: }
\begin{equation}\label{lambdaR}
	\lambda(R) = \lambda_0 \left(1 + \frac{R}{\mu} \right),
\end{equation}
{where $\mu$ is a physical scale with dimension of [Length]$^{-2}$, and $\lambda_0$ is the universal CSL parameter. In the cosmological setting,  if one wishes to avoid introducing fine-tuning aspects into $\lambda(R)$, there are essentially two natural options for $\mu$. The first option is $\mu = \ell_P^{-2}$, where $\ell_P$ is Planck's length. In this case, $R/\mu \ll 1$ because inflation involves energy scales less than the Planck scale. The second reasonable choice is $\mu = H^2$. Here, $R/\mu \simeq 12$ since $R \simeq 12 H^2$ during inflation. Consequently, in both cases, we find that $\lambda(R) \simeq  \lambda_0 \times \mathcal{O}(10)$. Based on this brief analysis, and for this particular CSL inflationary model, we are enabled to consider $\lambda \simeq \lambda_0$ during inflation, which implies that Eq. \eqref{parametrizacion0} is now expressed as }
\begin{equation}\label{parametrizacion} 
	\lambda_k = \lambda_0 k .
\end{equation}
{However, we acknowledge that the former expression relied on certain assumptions that are far from unique. }

{For the purposes of this work, with an aim to align with astronomical considerations, we will set the numerical value of the CSL parameter as $\lambda_0 = 10^{-14}$ s$^{-1}$, or equivalently, $1.029$ Mpc$^{-1}$ in more convenient units for our analysis.} This value is consistent  with empirical constraints obtained from experimental data, including spontaneous X-ray emission \citep{sandro2017}, matter-wave interferometry \citep{toros2016}, gravitational wave detectors \citep{carlesso2016}, and neutron stars \citep{cota_lambda0}.

After setting this reference for the functional dependence of $\lambda_k$, we have checked that estimations for cosmological parameters from the CSLIM, including the running of the spectral index,  are statistically consistent with the $\Lambda$CDM$+$running model (see Sec. \ref{Sec:ref_model}), providing a meaningful consistency check.  This was expected from the previous analysis in temperature auto-correlation \citep{running2020}. At this point, we can conclude that our model presents a novel explanation for the emergence of primordial inhomogeneities, and it fits the data remarkably well, in spite of not offering distinct predictions for observational comparison. 

However, there is no a priori reason to assume that the form of $\lambda_k$  shown in Eq. \eqref{parametrizacion} is the only possible one. Therefore, we propose new functional dependencies that introduce a small departure from the original one, characterized by a new free parameter $B$, these are:
\begin{subeqnarray}\label{eq:params12y3}
	\lambda_k &=& \lambda_0 (k+B), \quad [B] = \rm{Mpc}^{-1} \label{previous}\\
	\lambda_k &=& \lambda_0 (k+B/k), \quad [B] = \rm{Mpc}^{-2} \\
	\lambda_k &=& \lambda_0 (k+B \, k^2), \quad [B] = \rm{Mpc}.
\end{subeqnarray}
{These different forms of $\lambda_k$ will introduce
	distinctive features in the angular anisotropy power spectrum, which will
	be further analyzed in Sec. \ref{sec:pprimplots}. Furthermore, these  three different functions  of $k$ parameterized by $B$, correspond to three different versions or ``schemes''  of the CSLIM.  Consequently, throughout the rest of this article, we refer to each of them (from top to bottom) as scheme I,  II, and III respectively. }

In the next section, we will estimate cosmological parameters for each case and compare them with a reference model.

{Before concluding this subsection, we offer an alternative motivation for considering the three schemes of \eqref{eq:params12y3}. In Ref. \citep{pedrocsl}, it was demonstrated that the simplest functional dependence $\lambda_k =\lambda_0 k$ can indeed be interpreted as a specific form of the collapse operator. Specifically, when applying the inverse Fourier transform of the corresponding variables in the CSL evolution equation \eqref{CSLevolution}, the resulting CSL equation in configuration space is as follows:}

\begin{eqnarray}\label{CSLevolution2}
	|\Phi, \eta \ket &=& \hat T \exp \bigg\{ - \int_{\tau}^{\eta} 
	d\eta'   \bigg[ i \hat{H} \nn
	&+& \frac{1}{4 \lambda_0} \int d^3x (w(\x,\eta') - 2 \lambda_0
	\hat{Y}(\x))^2 \bigg] \bigg\} |\Phi, \tau \ket, \nn
\end{eqnarray}
{which is just the standard CSL evolution state vector with the universal parameter $\lambda_0$, and an effective collapse generating operator $\hat Y$ defined as $\hat{Y}(\x) \equiv (-\nabla^2)^{1/4} \hat y (\x)$.}

{In Ref. \citep{pedrocsl}, it was also argued that no satisfactory answer exists for explaining the fundamental reason determining the appearance of the operator $(-\nabla^2)^{1/4} \hat y (\x)$. Ultimately, a sound justification will have to wait for a general theory that expresses, in all situations, from particle physics to cosmology, the exact form of the CSL-type modification to the evolution of quantum states.  Therefore, the schemes proposed in \eqref{eq:params12y3} can be interpreted as the next leading order modifications to the operator $(-\nabla^2)^{1/4} \hat y (\x)$. In particular, we are exploring possible corrections of the effective collapse operator by expanding $\hat{Y}(\x)$ as a ``Taylor series'' at the next leading order in $\nabla^2$. For example, scheme III can be interpreted as the effective collapse operator   
	\begin{equation}
		\hat{Y}(\x) = [(-\nabla^2)^{1/4} + (B/2) (-\nabla^2)^{3/4} ] \hat y(\x) ,
	\end{equation}	
	provided $B$ small.
	Similarly, for schemes I and II, we are considering the next leading order terms of the collapse operator $\hat{Y}(\x)$, which are of the form  $(B/2) (-\nabla^2)^{-1/4} \hat y(\x)$ and $(B/2) (-\nabla^2)^{-3/4} \hat y(\x)$ respectively. }



\section{Observational Constraints}\label{Sec3}
In this section we will constrain the $\Lambda$CDM parameters together with the free parameter $B$ introduced in \eqref{eq:params12y3}, corresponding to the CSL inflationary model.

We proceeded to modify the {\sc camb} \citep{CAMB} source code\footnote{Original code is available at \href{https://github.com/cmbant/CAMB}{https://github.com/cmbant/CAMB}.} to include our calculations in the primordial power spectrum. Afterwards, the modifications were incorporated to the {\sc cosmomc} \citep{COSMOMC} (which was  adapted to include the new parameter of interest) an then analyzed with {\sc getdist} codes\footnote{Original code can be obtained at \href{https://github.com/cmbant/CosmoMC}{https://github.com/cmbant/CosmoMC}.}.

Observational data used corresponds to the latest \citep{Planck18like} data consisting in {\em Planck} temperature and polarization correlations likelihood {\em plikTT,TE,EE+lowE+lensing}\footnote{Code available at \href{https://pla.esac.esa.int}{https://pla.esac.esa.int}.}.  {The latter name follows the CMB likelihood naming convention adopted by the \textit{Planck} papers\footnote{  {The information content of the CMB sky can be split into temperature ($T$), plus two polarization components, the $E$ and $B$ modes} } \citep{Planck18like}: \textit{Planck TT} labels the likelihood formed using only the temperature data, spanning the multipole range $2 \leq \ell \leq 2500$; \textit{Plank TE} and \textit{Planck EE} labels the likelihood formed using exclusively the \textit{TE} power spectrum from $30 \leq \ell \leq 2000 $ and the \textit{EE} power spectrum respectively. Henceforth \textit{plik TT,TE,EE} labels the combination of \textit{Planck TT, Planck TE} and \textit{Planck EE}, taking into account correlations between the \textit{TT, TE, EE} spectra at $\ell >29$. Additionally \textit{lowE} labels the likelihood formed using the \textit{EE} power spectrum over $2 \leq \ell \leq 30$ and \textit{lensing} labels the lensing likelihood which includes lensing effects.}

\subsection{Primordial Power Spectrum}

Employing the CSL inflationary model described in Sec. \ref{Sec2}, in Ref. \citep{running2020} we have calculated the scalar power spectrum up to second order in the HFF, which is given by
\begin{equation}\label{pdek}
	\mP_s(k) = A_s \left( \frac{k}{k_\diamond} \right)^{n_s-1 + \frac{\nrun}{2} \ln \frac{k}{k_\diamond}       } C(k),
\end{equation}
where
\begin{equation}
	A_s \equiv \frac{ H_\diamond^2}{ \pi^2 M_P^2 \epsilon_{1\diamond}}.
\end{equation}
{In our notation, $k_\diamond$ represents a pivot scale, defined as $k_\diamond |\eta_\diamond| = 1$, indicating that the mode has ``crossed the horizon'' at the time $|\eta_\diamond|$. Therefore,  $H_\diamond$ and $\epsilon_{i\diamond}$ denotes that these variables are being evaluated at the time of ``horizon crossing'' $|\eta_\diamond|$. We will fix the pivot scale $k_\diamond$  at the usual value of $0.05$   $\mathrm{Mpc}^{-1}$. }

Furthermore, $n_s$ represents the scalar spectral index and $\nrun$ the running of the spectral index. Theoretically, our prediction for those parameters at the lowest order in the HFF is: $n_s = -2 \eidi - \eiidi$ and $\nrun = -2 \eidi \eiidi - \eiidi \epsilon_{3 \di}$.

The function $C(k)$ contains the effects of the CSL inflationary model, and is expressed in terms of the aforementioned parameters.  {In Ref. \citep{running2020}, we have calculated $C(k)$ in detail up to the second order in the HFF. Subsequently, we have expressed the HFF as a function of the corresponding inflationary parameters, resulting in the following equation:}
\begin{eqnarray}\label{defCK2}
	&C(k)& = 1 + \frac{\lambda_k   |k \tau|}{ k^2} + \frac{\lambda_k }{2 k^2} \cos(2 |k \tau|)   \nn 
	&-&   \exp \bigg\{   \bigg[ -4+ n_s + \nrun  \ln \frac{2 k}{k_\diamond}    \bigg]  \ln \zeta_k - \frac{\nrun}{2}  \left( \ln^2 \zeta_k - \theta_k^2  \right)   \bigg\}  \nn
	&\times& \bigg[ \cos \bigg\{ \left[  -4 +n_s + \nrun \ln \frac{2 k}{k_\diamond}  \right] \theta_k - \nrun  \left(  -\Delta N_\diamond + 1 -\ln 3 + \ln \frac{k}{2 k_\diamond}  \right) \nn
	&\times& \theta_k \ln \zeta_k  \bigg\} \bigg]^{-1}.
\end{eqnarray}
{In this expression $\Delta N_\diamond$ is the number of e-foldings from the horizon crossing of the pivot scale to the end of inflation; for the pivot scale $k_\diamond = 0.05$ Mpc$^{-1}$, this number corresponds to $\Delta N_\diamond \simeq 55$ \cite{Planck18b}. } We require that the localization (collapse) process is fast enough compared to the total duration of inflation, in conformal time, this condition is expressed as $\lambda_0 |\tau| > 1$.   {In the following, we will provide a brief analysis of the calculation of $|\tau|$. }

{We begin by invoking the definition of the total number of e-foldings during inflation, denoted as $N_\fin$, this is $a_\fin \equiv e^{N_\fin} a(\tau)$. Next, we substitute the functional form of the scale factor during inflation $a(\eta)\simeq-1/(H \eta)$ (with $H \simeq$ const.) in the right hand side of the former expression, obtaining $|\tau| = e^{N_\fin}/(a_\fin H_\fin)$. Additionally, by employing Friedmann's equation, $H^2_\fin = \rho_\fin/(3 M_P^2)$, we can determine that}
\begin{equation}\label{tauexp}
	|\tau| \simeq \frac{e^{N_\fin} \sqrt{3} M_P}{a_\fin \sqrt{\rho_\fin}}.
\end{equation}
{To estimate $a_\fin$,  we make the simplification that the reheating era is instantaneous. Also, by taking into account that for pure radiation, the corresponding energy density satisfies $\rho_{\textrm{rad}} a^4=$ constant, we can write $a_\fin/a_0 = (\rho_0/\rho_\fin)^{1/4}$. At the present epoch, radiation primarily comprises CMB photons, which exhibit a blackbody radiation spectrum. Hence, $\rho_0 = (\pi^2/15) T_0^4$, where $T_0$ represents the current temperature of the CMB. By normalizing the scale factor today $a_0 = 1$,  and using the latter expression for $\rho_0$,  we obtain the following equation:}
\begin{equation}\label{afin}
	a_\fin = \left( \frac{\pi^2}{15 \rho_\fin} \right)^{1/4} T_0.
\end{equation}
{Substituting \eqref{afin} into \eqref{tauexp}, yields}
\begin{equation}\label{tauexp2}
	|\tau| \simeq e^{N_\fin}\sqrt{3} M_P \left( \frac{15}{\pi^2 \rho_\fin} \right)^{1/4} \frac{1}{T_0}.
\end{equation}
{Here, we will make the following assumptions: inflation ends at an energy scale $\rho_\fin^{1/4} \simeq  10^{-3} M_P$; and  the total duration of inflation is\footnote{The approximated minimum number of e-foldings to solve the horizon  problem is $\exp[ N_{\textrm{min}} ] \simeq \rho_\fin^{1/4}/(0.037 h$ eV$)$ \citep{Weinberg2008}. Consequently, taking $h \simeq 0.7$ and $\rho_\fin^{1/4} \simeq  10^{-3} M_P$, implies $N_{\textrm{min}} \simeq 60$. Thus, in our model we assume 7 e-foldings more than this minimum for the total duration of inflation. } $N_\fin=67$.  Substituting these  numerical values of $\rho_\fin$ and $N_\fin$ in Eq. \eqref{tauexp2}, and taking into account that the most recent data \citep{Planck18a} suggest $T_0 \simeq 2.4 \times 10^{-13}$ GeV, we ultimately obtain the value $|\tau| = 8 \times 10^{6}$ Mpc by making the appropriate units conversions; also we can check that  $\lambda_0 |\tau| >1$.}

The quantities $\theta_k$ and $\zeta_k$ in Eq. \eqref{defCK2} are defined as:
\begin{equation}\label{defzetakythetak_main}
	\zeta_k \equiv \left( 1 + \frac{4 \lambda_k^2}{k^4}   \right)^{1/4}, \qquad \theta_k \equiv - \frac{1}{2} \arctan \left( \frac{2 \lambda_k}{k^2} \right).
\end{equation}
A few lines above we mentioned that $C(k)$ encodes the effects of the CSL mechanism. As a matter of fact, we note that if $\lambda_k=0$, which means no collapse, then $\zeta_k=1$ and $\theta_k =0$, giving $C(k)=0$. That is to say, if there is no collapse
of the wave function, then the vacuum state remains unchanged, no perturbations
are generated and hence $\mP(k)=0$ at all scales; in our model this is what we expected.

The main modification that needs to be incorporated into the {\sc camb} code is given by Eq. \eqref{pdek}. This modification solely impacts the inflationary period;  {however, the behavior of such a complex equation [see Eq. \eqref{defCK2}] must be carefully monitored before migrating into numerical calculations. When transitioning from analytical expressions to a programming language (such as {\sc camb}), it is crucial to ensure that the effects observed in the results are attributed to the theoretical model itself and not to numerical representation issues.}

{To begin our analysis, we implemented the expression for the primordial power spectrum in the {\tt Fortran} language, which is the same programing language used for {\sc camb} and {\sc cosmomc}. We made appropriate adjustments to the equation to minimize numerical errors. During execution, no ``underflows'' or ``overflows'' were reported \citep{Goldberg91}. Furthermore, we conducted various consistency tests. At first, we ensured that the standard $\Lambda$CDM primordial power spectrum shape was recovered in the appropriate conditions (i.e. $B=0$). 
	Afterwards, we progressively explored a range of feasible values for the free parameter $B$. In all cases, no errors arose, and the behavior of these tests met our expectations;  there were no unexpected features, such as jumps or discrepant values. These steps are crucial for instilling confidence in future results, as more complex codes may inadvertently mask pre-existing numerical issues. We followed a similar procedure when incorporating the expression \eqref{pdek} into the {\sc camb} code to ensure that the results obtained through {\sc cosmomc} are reliable.}

	%
	%
	%
	%

\subsection{Reference model}\label{Sec:ref_model}



As mentioned earlier, the simplest functional form $\lambda_k = \lambda_0 k$  can be comparable to the standard cosmological model. Therefore, our first consistency check involves contrasting parameter estimations between the $\Lambda$CDM model\footnote{This model was built with the same datasets and using an unmodified {\sc camb} plus {\sc cosmomc} code. The reason to run our own $\Lambda$CDM is to assure the same input parameter configuration for all the models, so that they could be compared on an equal footing.} and the CSLIM with $B=0$, i.e. corresponding to $\lambda_k = \lambda_0 k$. Table~\ref{tab:refmodel} shows the estimations at a $68\%$ confidence level for the reference model parameters plus $n_{\rm run}$ and the CSLIM estimations. Both models demonstrate statistically identical results.

	%

Regarding observational constraints, both models yield the same parameter estimation and are equivalent in this respect, with no distinctive features that set them apart. However, as we mentioned in previous sections, there is no reason to favor this specific form of $\lambda_k$, so we can proceed to test other three schemes.

Henceforth, we will refer to this scenario as the ``reference model'', characterized by the standard $\Lambda$CDM cosmological model, which will serve as a baseline for depicting the distinct features of the CSLIM.

\begin{table}
	\centering
	
	\caption{Comparison of parameter estimation between the standard cosmological model and the CSLIM with $B=0$, i.e.  $\lambda_k = \lambda_0 k$. From an observational point of view, they are indistinguishable. The difference lies in the theoretical explanation of the origin of the primordial cosmic perturbations.  }
	\label{tab:refmodel}
	\setstretch {1.5}
	\begin{tabular} { l  c  c}
		Parameter & $\Lambda$CDM  &  CSLIM \\
		\hline
		{\boldmath$\Omega_b h^2$} & $0.02239\pm 0.00015$  & $0.02239\pm 0.00015 $\\
		{\boldmath$\Omega_c h^2$} & $0.1203\pm 0.0014$  & $0.1204\pm 0.0014 $\\
		{\boldmath$100\theta_{MC} $} & $1.04089\pm 0.00031$  & $1.04090\pm 0.00031$\\
		{\boldmath$\tau_{\rm{d}} $} & $0.0559^{+0.0075}_{-0.0083}$ & $0.0559\pm 0.0082  $\\
		{\boldmath${\rm{ln}}(10^{10} A_s)$} & $3.050^{+0.016}_{-0.018}$ & $3.050\pm 0.017 $\\
		{\boldmath$n_s $} & $0.9636\pm 0.0047$  & $0.9635\pm 0.0047        $\\
		{\boldmath$n_{\rm run}$} & $-0.0059\pm 0.0067$   & $-0.0059\pm 0.0068  $\\
		\hline
	\end{tabular}
\end{table}

\subsection{Primordial Power Spectrum Analysis}\label{sec:pprimplots}

In the previous subsection we have established the {re\-fe\-rence} model, i.e. by assuming $\lambda_k = \lambda_0 k$, we have found that the CSLIM  
reproduces the standard predictions of inflation plus the $\Lambda$CDM model, see Table \ref{tab:refmodel}. In the present subsection we move forward and include the $B$ parameter in our analysis. 

We note first that the CSL mechanism introduces native oscillatory features in the scalar power spectrum $\mP_s (k)$ at low values of $k$ (large angular scales); this effect is caused by the third term in \eqref{defCK2}, i.e. by the cosine function. It is important to notice that these oscillatory effects cannot be completely ``turned off'' even when $\lambda_k = \lambda_0 k$. In fact, these oscillations reflect the characteristics of the initial quantum state, specifically the Bunch-Davies vacuum at time $\tau$ --represented as plane waves $\sim e^{-i k |\tau|}/\sqrt{2k}$-- which subsequently evolves in accordance with CSL dynamics (see Appendix A of Ref. \cite{running2020}).

Next, we focus on the three schemes outlined in \eqref{eq:params12y3}. These schemes manifest distinct characteristics in the primordial power spectrum due to their varying functional dependencies on $k$. Our analysis will primarily concentrate on the observational range, which is $10^{-6} \mathrm {Mpc}^{-1} \leq k \leq 10^{-1} \mathrm{Mpc}^{-1}$.

\begin{figure}[h]
	\centering
	\includegraphics[width=0.55\textwidth,angle=270]{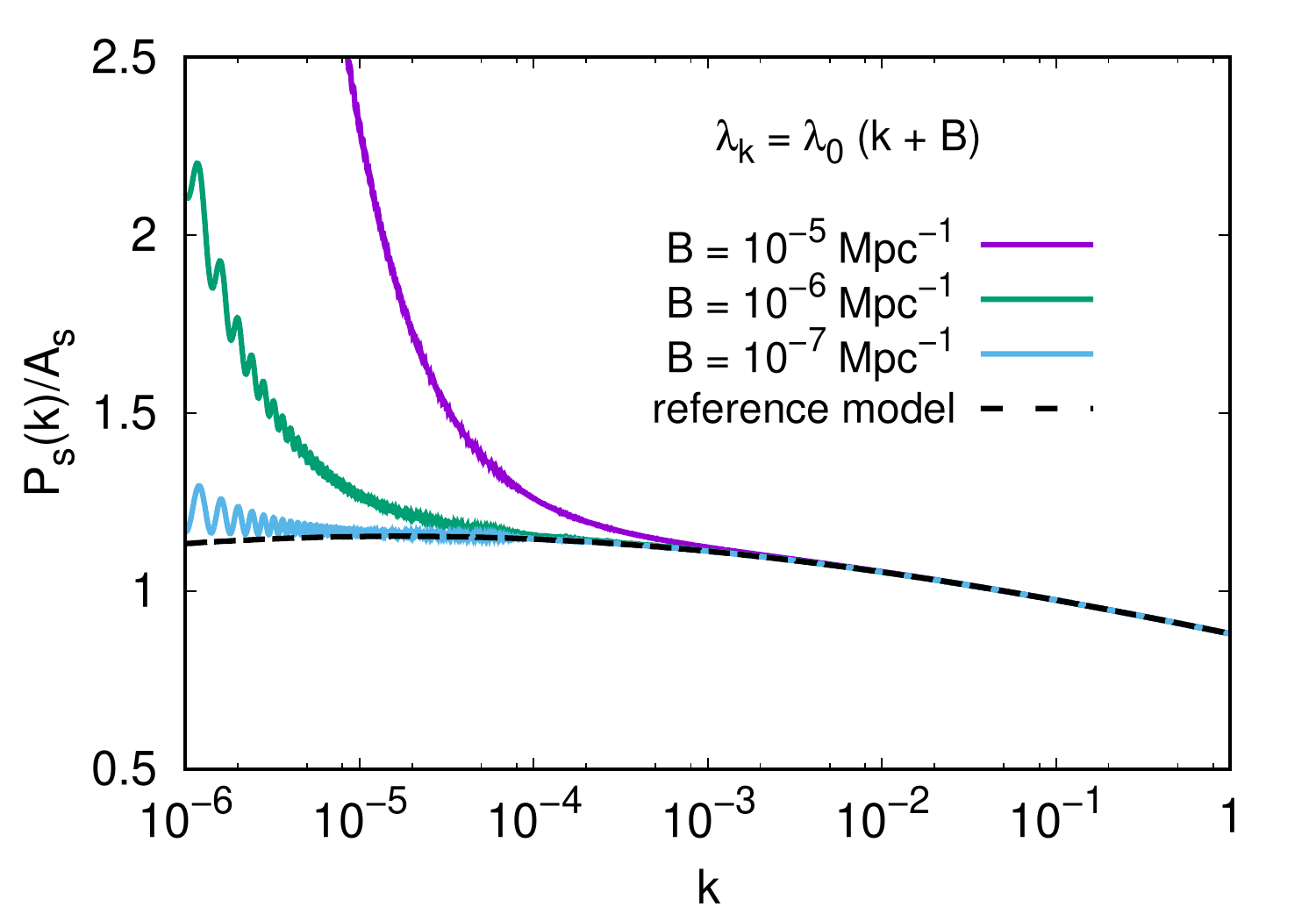}
	\caption{The predicted power spectrum for scheme I, normalized by its amplitude $A_s$. The plot illustrates the effect of varying $B$  in the interval $10^{-6} \mathrm {Mpc}^{-1} \leq k \leq 10^{-1} \mathrm{Mpc}^{-1}$. As the value of $B$ increases, a progressive departure from the reference model is observed for lower $k$.}
	\label{fig:PdeKparam0}
\end{figure}

In the first scheme, as depicted in Fig. \ref{fig:PdeKparam0}, it can be observed that as the value of $B$ decreases, the shape of $\mP_s(k)$ in the CSL model increasingly resembles that of the reference model. Some oscillations still persist for $k < 10^{-4}$ Mpc$^{-1}$, but they do not significantly affect the spectrum at higher values of $k$.

\begin{figure}[h]
	\centering
	\includegraphics[width=0.55\textwidth,angle=270]{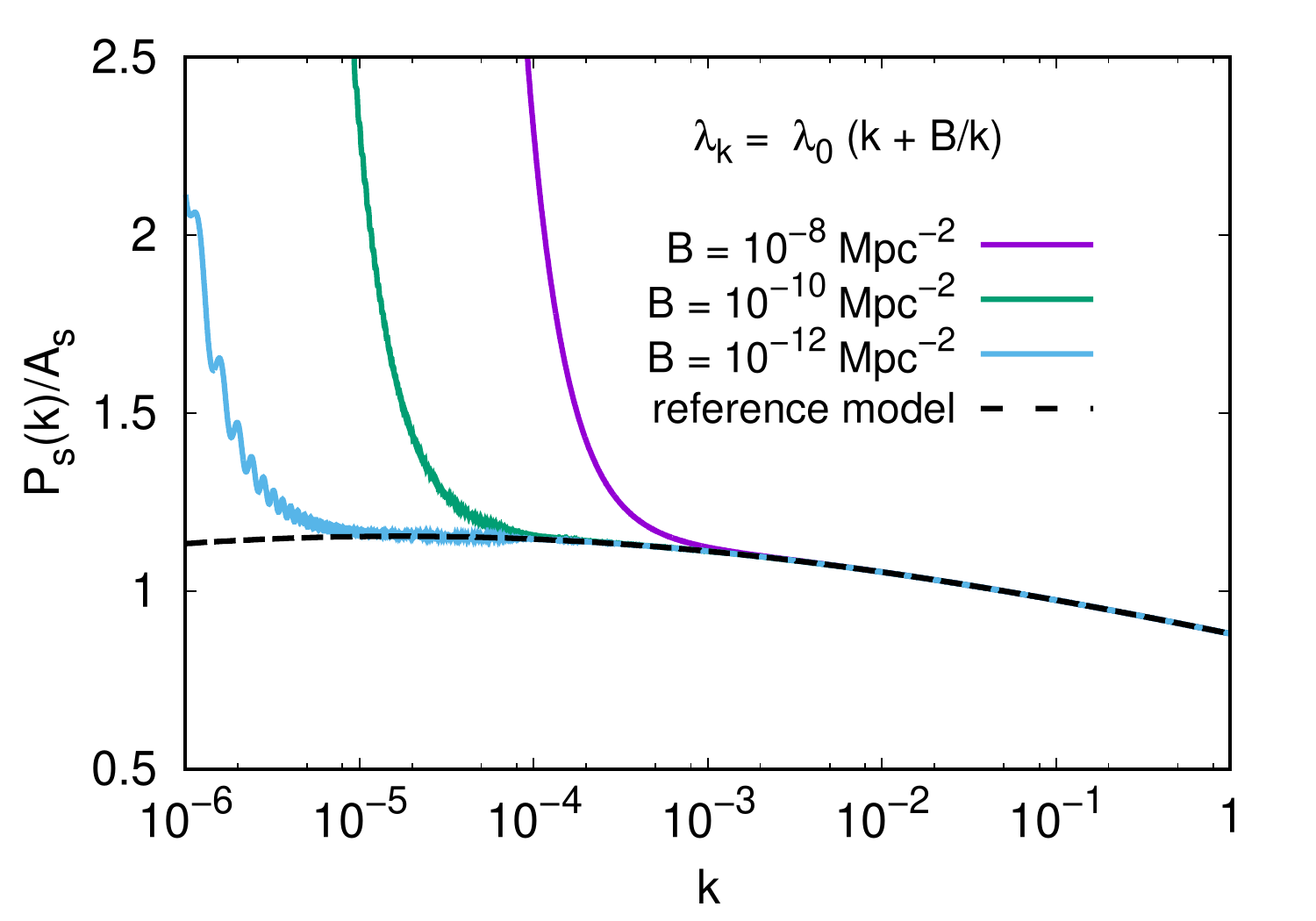}
	\caption{The predicted power spectrum for scheme II, normalized by its amplitude $A_s$. We observe a similar behavior to scheme I, but in this case the spectrum is more sensitive to the value adopted for $B$. }
	\label{fig:PdeKparam1}
\end{figure}

A similar behavior is found for the second scheme. The predicted power spectrum, normalized by its amplitude $A_s$, is plotted
in Fig. \ref{fig:PdeKparam1}. In this case, smaller values of $B$ are required to better match the shape of the reference model. However, it is important to emphasize that the primary aim of the CSLIM is not to precisely replicate the power spectrum of the standard inflationary model,  but rather to fit the observed data while preserving unique characteristics that might be able to distinguish it from the reference model (in particular the features attributed to the running of the spectral index). In the range from $B = 10^{-12}$ to $10^{-8}\, \rm{Mpc}^{-2}$, the plotted examples demonstrate that the effect of varying $B$ can significantly affect the spectrum at $k \leq 10^{-1} k_\diamond$.

\begin{figure}[h]
	\centering
	\includegraphics[width=0.55\textwidth,angle=270]{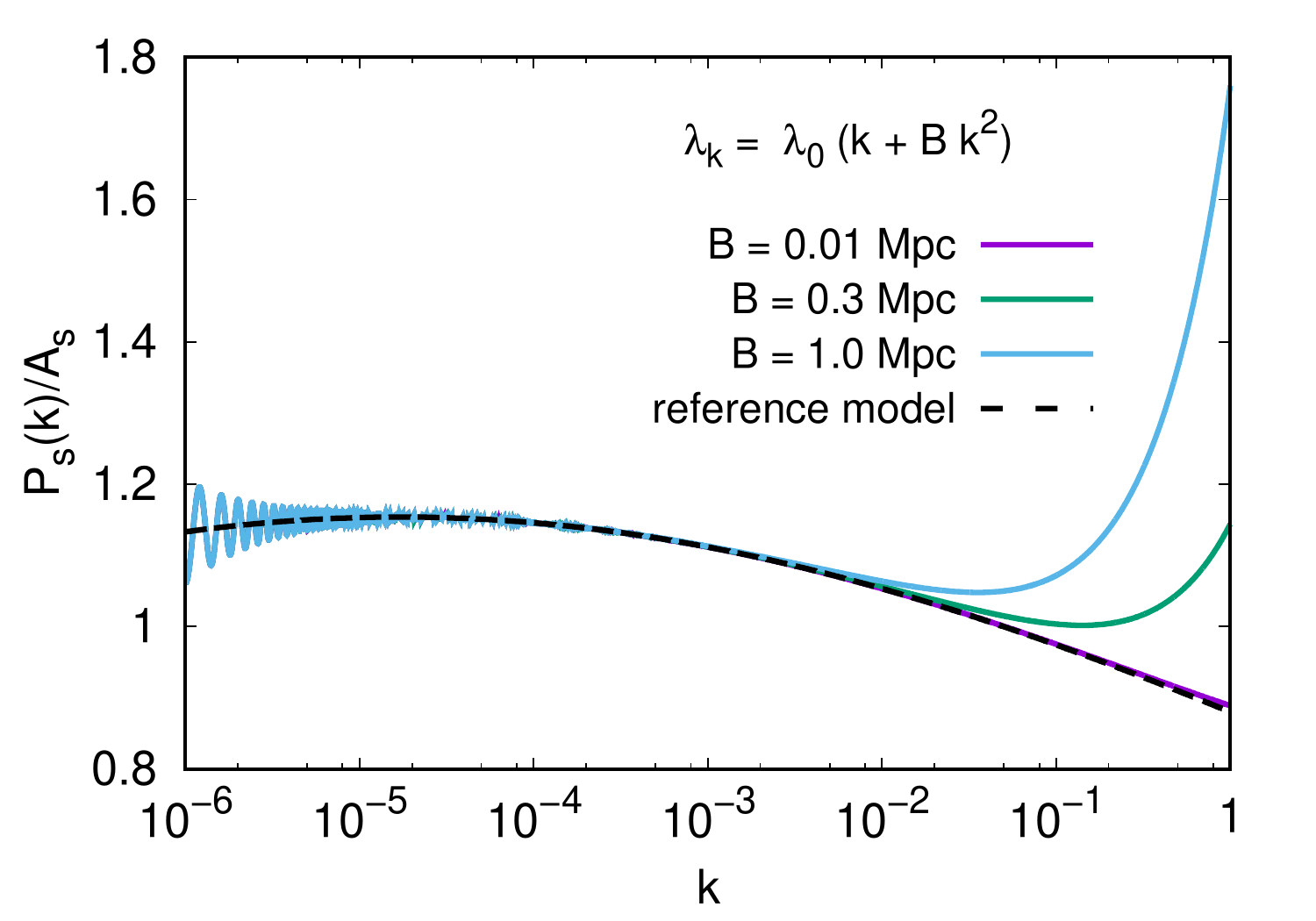}
	\caption{The predicted power spectrum for scheme III, normalized by its amplitude $A_s$. The functional shape proposed for scheme III primarily impacts large values of $k$. }
	\label{fig:PdeKparam2}
\end{figure}

Similar to the previous cases, oscillations remain prominent at low values of $k$ in the third scheme. However, due to the quadratic dependence on $k$, the influence of  $B$ becomes noticeable at larger values of $k$ (small angular scales), see Fig. \ref{fig:PdeKparam2}.  The characteristic oscillations persist at small values of $k$ (large angular scales) in this case. Exploring values of $B$ ranging from $10^{-2}$ Mpc to $1$ Mpc, we observe that as $B$ increases, the deviation of the CSLIM primordial power spectrum from the shape of the fiducial model becomes more pronounced.

At this point of the analysis, we have identified the distinctive features that the CSLIM introduces in the primordial power spectrum $P_s(k)$, and examinated the effect of varying the free parameter $B$ in the three schemes.  For the second part of our analysis,  we  proceed to investigate whether these effects can in fact be observed. To achieve this task, we focus on the CMB anisotropies angular power spectrum, which is  one of the main observables at hand.  We will only present the analysis of  the temperature autocorrelation (TT) spectrum, as the E-mode
autocorrelation (EE) and the {tem\-pe\-ra\-tu\-re}--E-mode cross-correlation (TE) spectra exhibit
the similar behavior.

\begin{figure}[h]
	\hspace{-1em}\includegraphics[width=0.55\textwidth,angle=270]{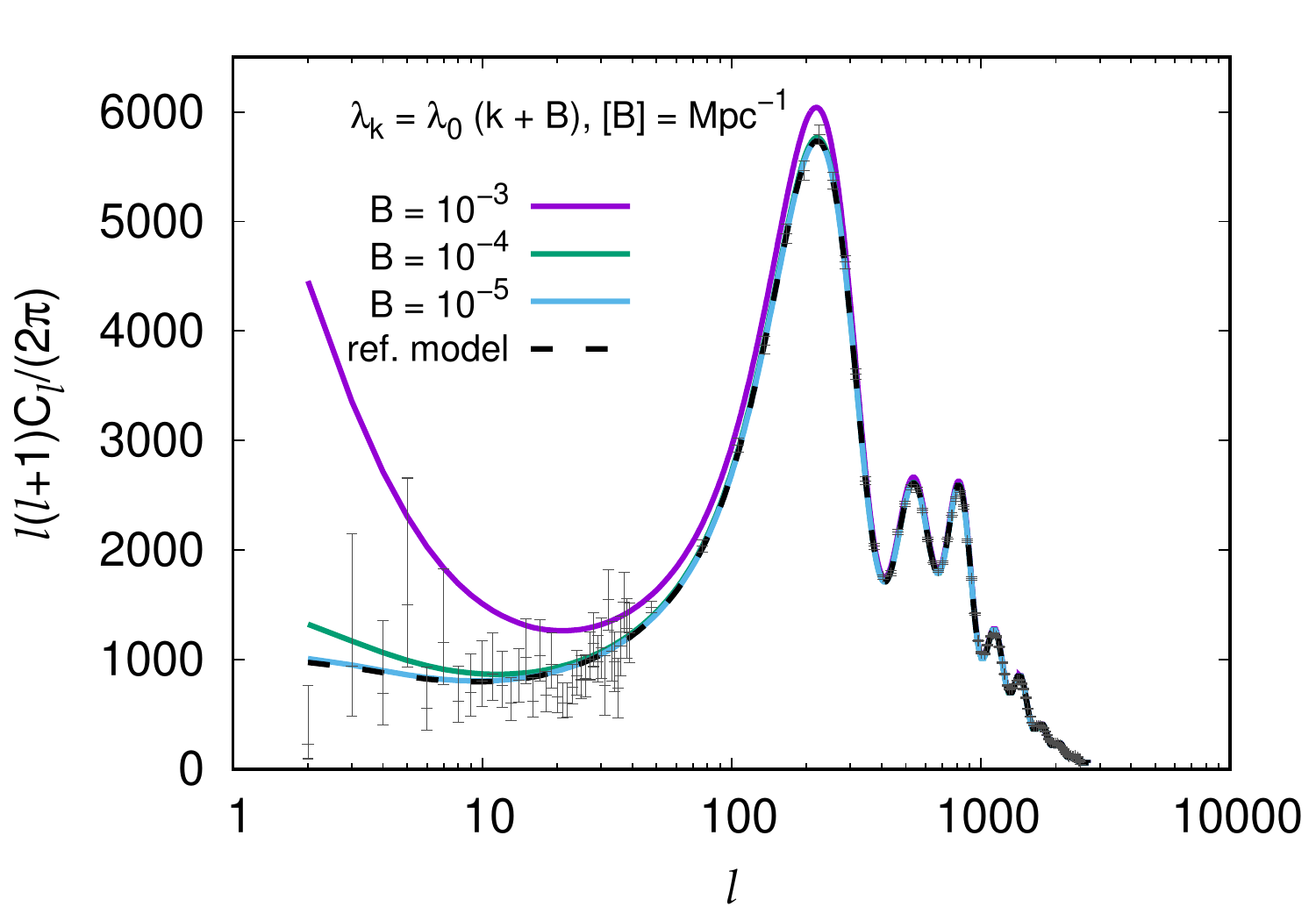}
	\caption{The CMB TT anisotropy angular spectrum is presented for scheme I, together with the reference model (dashed line) and
		{\em Planck} data points (with 1-$\sigma$ error bars) in gray. Different values of $B$ are considered. 
		For $B \simeq 10^{-3}$ Mpc$^{-1}$, the predicted spectrum at low multipoles and the height of its first peak deviate significantly from the
		standard one. For values of the free parameter on the
		order $10^{-5}$ Mpc$^{-1}$, the CSLIM spectrum completely matches the reference model. An intermediate value
		can be estimated to match observational data while preserving the unique features introduced by the CSLIM.}
	\label{fig:Clsparam0}
\end{figure}

\begin{figure}[h]
	\hspace{-1em}\includegraphics[width=0.55\textwidth,angle=270]{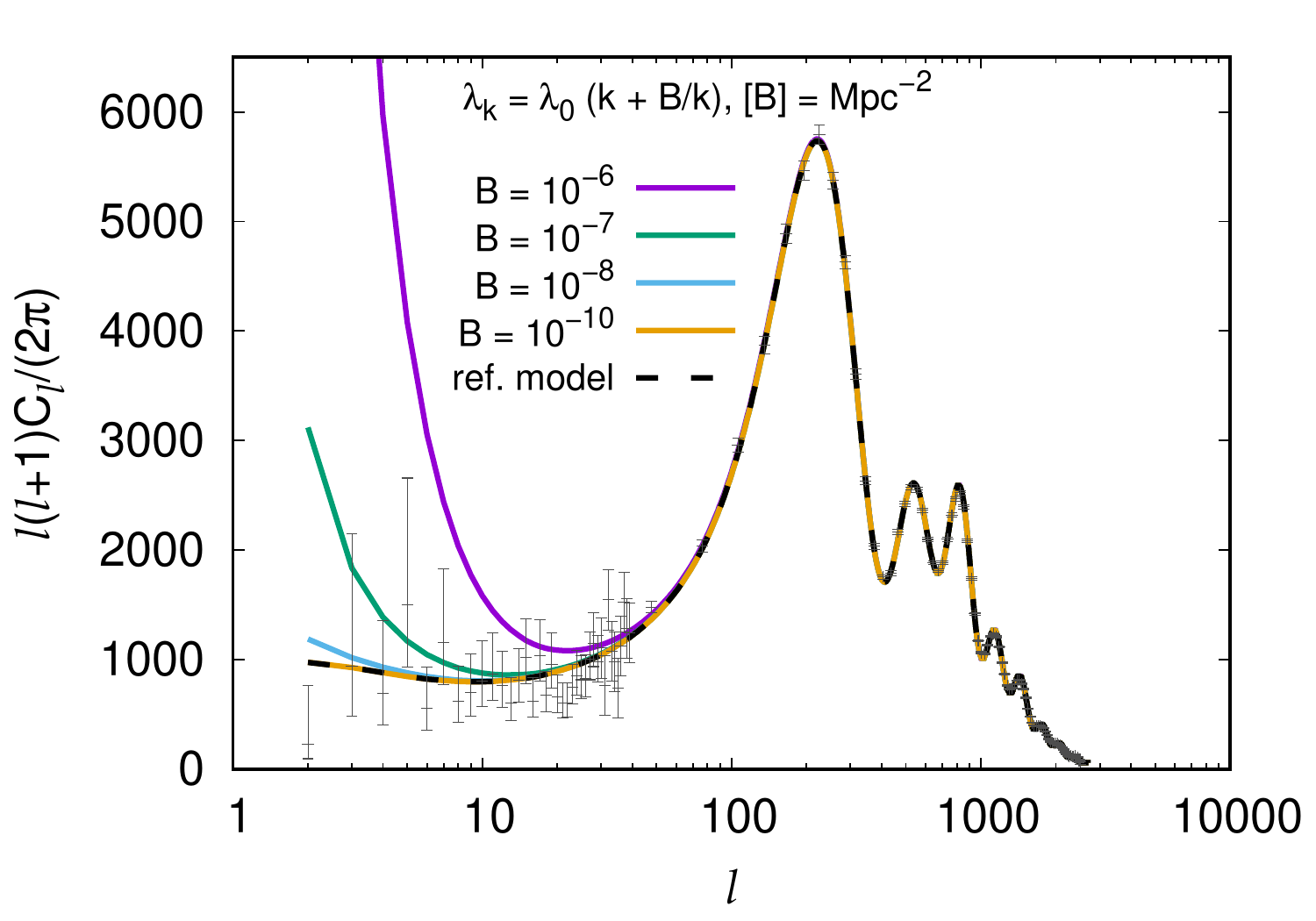}
	\caption{The CMB TT anisotropy angular spectrum is presented for scheme II, together with the reference model (dashed line) and
		{\em Planck} data points (with 1-$\sigma$ error bars) in gray. In this case, 
		significantly lower orders of magnitude for $B$ are required to match the reference model.
		For the values used in this example, a noticeable deviation is observed for $l < 40$. Although the variation of parameter $B$ in this scheme primarily influences the range in the spectrum that is measured with great uncertainty, particularly within the cosmic variance region, it is still feasible to discern an impractical range of values for the free parameter.}
	\label{fig:Clsparam2}
\end{figure}

\begin{figure}[h]
	\hspace{-1em}\includegraphics[width=0.55\textwidth,angle=270]{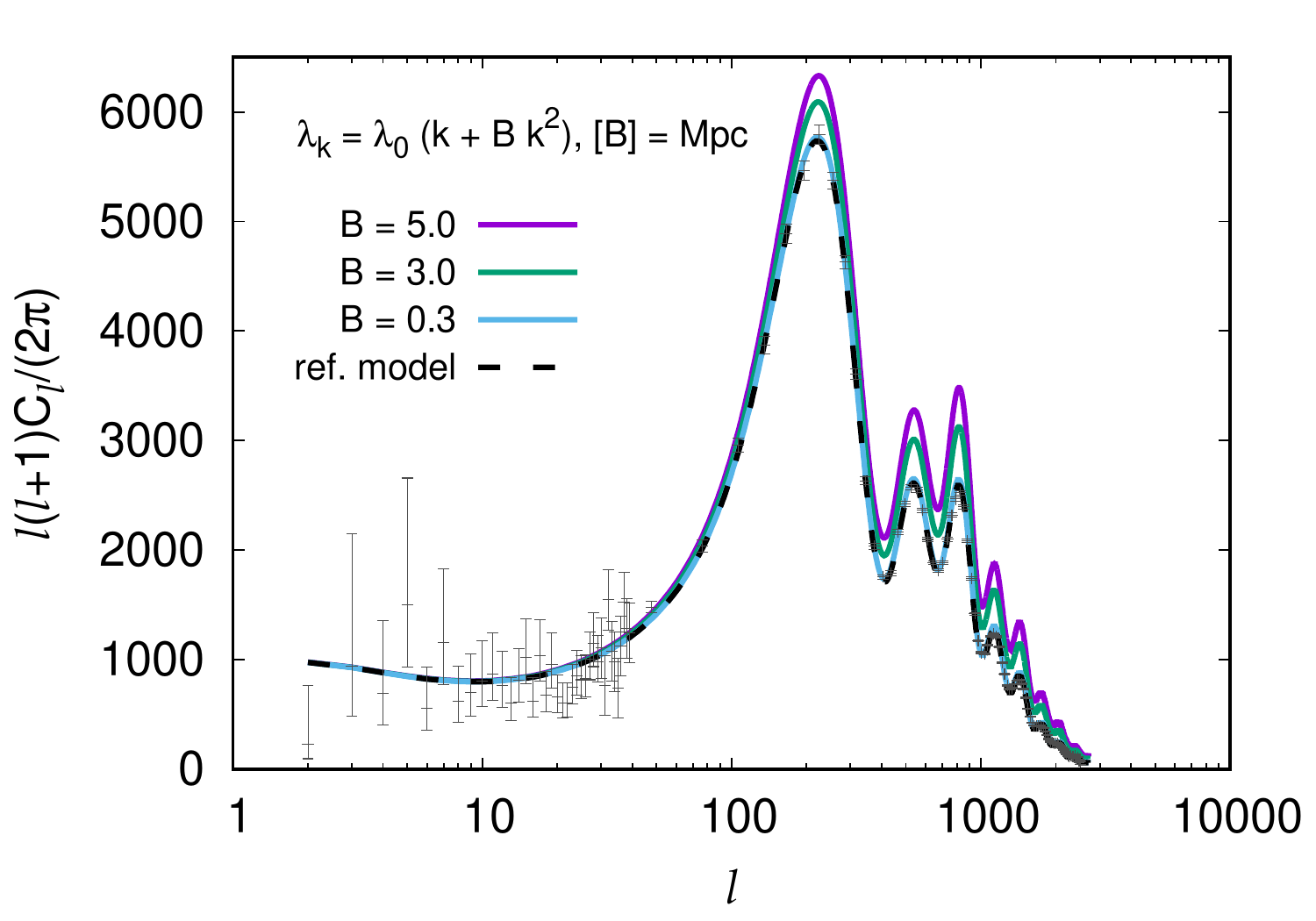}
	\caption{The CMB TT anisotropy angular spectrum is presented for scheme III, together with the reference model (dashed line) and
		{\em Planck} data points (with 1-$\sigma$ error bars) in gray. Similar to the other two schemes, the oscillatory features that were initially present in the primordial power spectrum are lost. However, the deviation from the reference model at large $l$ becomes crucial when constraining the value of  $B$ with observational data. }
	\label{fig:Clsparam3}
\end{figure}

As expected from the analysis of Figs. \ref{fig:PdeKparam0} and  \ref{fig:PdeKparam1}, schemes I and II display progressive departures from the
reference model as the value of $B$ increases, see. Figs. \ref{fig:Clsparam0}
and \ref{fig:Clsparam2}. This effect occurs for large angular scales
(low values of $l$). In contrast, in scheme III the effect is seen at small angular scales
(large values of $l$), see Fig. \ref{fig:Clsparam3}, which is also consistent with the behavior displayed in Fig. \ref{fig:PdeKparam2}.

The oscillations observed in the primordial power spectrum for the three schemes, were located at low values of $k$. Consequently, we would expect these oscillations to be located at low values of $l$ in the corresponding angular power spectrum, i.e. within the  cosmic variance region. However,  from Figs. \ref{fig:Clsparam0}, \ref{fig:Clsparam2} and \ref{fig:Clsparam3}, it is clear that the oscillatory features are largely attenuated, but the net departure persists as discussed above.

Furthermore, it is evident that in schemes I and III (as depicted in Figs. \ref{fig:Clsparam0} and \ref{fig:Clsparam3}), the first peak of the spectrum, which position has been meticulously measured, is not reproduced for specific values of the free parameter. This distinction accentuates the capability of our model to depart from the standard model, making parameter $B$ suceptible to be constrained by observational data.
In summary, we can affirm that the novel features introduced by the CSLIM model are indeed observable.

\subsection{Estimation of $B$ parameter}
We now propose to statistically estimate the value of the $B$ parameter, introduced previously in Eq. \eqref{eq:params12y3}, in order to test whether the
CSLIM model has the potential to raise any observational feature that distinguishes it from the $\Lambda$CDM model.

\begin{figure}
	\centering
	\includegraphics[width=\columnwidth]{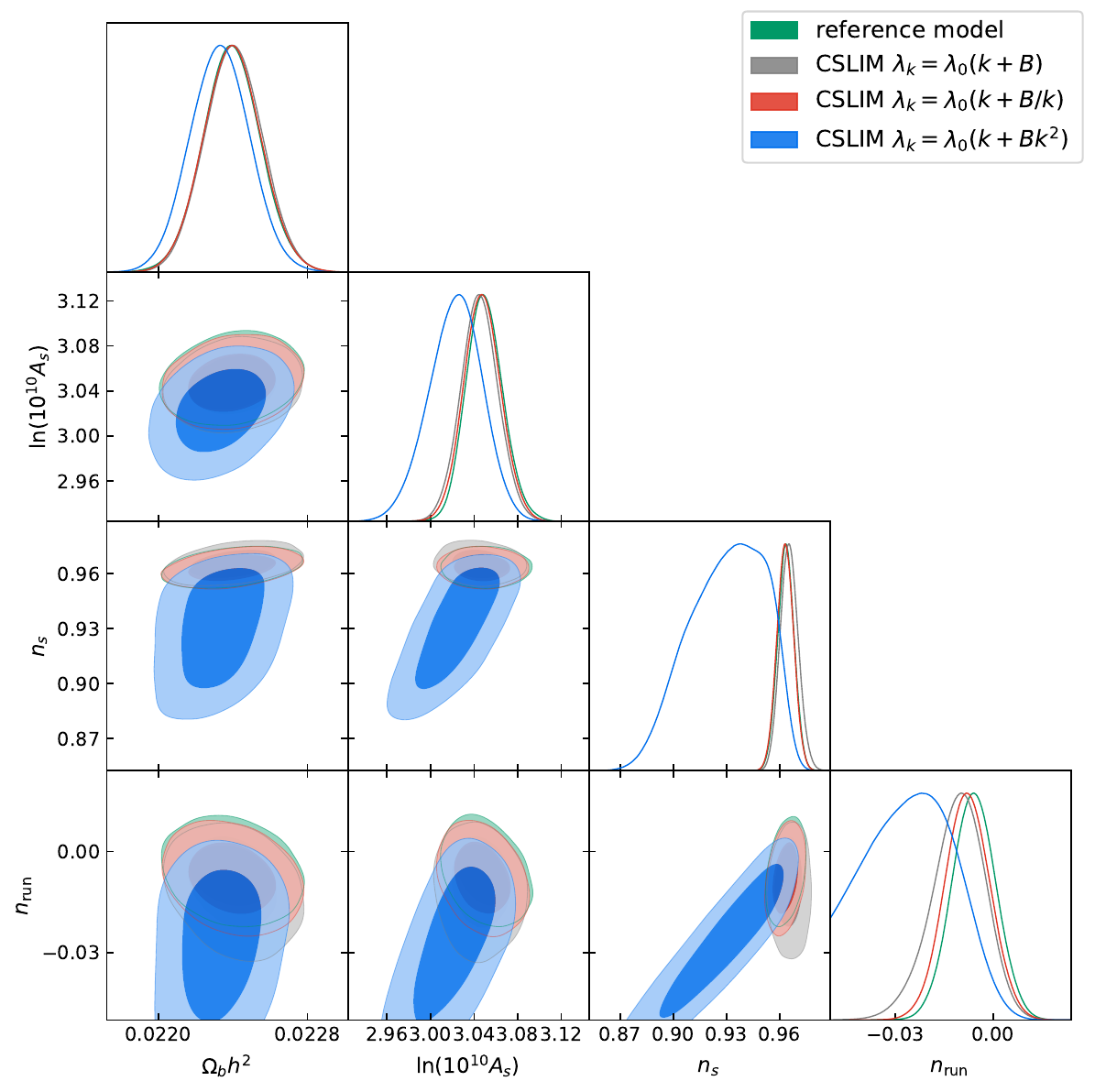}
	\caption{Comparison between parameter estimation in the reference model
		and the CSLIM using the three schemes in Eq. \eqref{eq:params12y3}. The $68\%$ and $95\%$ confidence level contours are plotted together with the posterior probability.  There
		is a difference in $n_s$, $A_s$ and $n_{\rm run}$ between the CSLIM and the reference model, the rest of
		the cosmological parameters yield the same results as the reference model.}
	\label{fig:planck_running-vs-paramsconrun}
\end{figure}

Figure~\ref{fig:planck_running-vs-paramsconrun} shows how the CSLIM, with the $B$ parameter, affects the scalar spectral index $n_s$, the amplitude ${\rm{ln}}(10^{10} A_s)$ and $n_{\rm run}$, which actually are the most affected parameters. One of the schemes introduces also a slight deviation for $\Omega_b h^2$.
In addition, the absolute magnitude of the running of the spectral index $|n_{\rm run}|$ changes with respect to the reference model. Table~\ref{tab:Bcero-planck_running} depicts these differences quantitatively, and also shows the results of the parameter estimation for $B$.
The rest of the cosmological parameters yield similar results to the reference model. 


As seen in Fig.~\ref{fig:planck_running-vs-paramsconrun}, the parameter space is constrained in various ways. Notably, different $k$ dependencies of $\lambda_k$ can lead to correlations between parameters. For instance, in the case where $\lambda_k = \lambda_0 (k+B k^2)$,  we observe correlations between $n_s$ and $n_{\rm run}$.


\begin{table*}
	\centering
	{\setstretch {1.7}
		\begin{tabular}{lcccc}
			Parameter &  $\Lambda$CDM & CSLIM I & CSLIM II & CSLIM III \\
			\hline
			{\boldmath${\rm{ln}}(10^{10} A_s)$} & $3.050^{+0.016}_{-0.018}$   & $3.045\pm 0.017$ & $3.047\pm 0.017$ &  $3.023^{+0.026}_{-0.023}$ \\
			{\boldmath$n_s  $} & $0.9636\pm 0.0047 $  &$0.9655\pm 0.0051$  &  $0.9633\pm 0.0047$  &  $0.931^{+0.027}_{-0.016} $ \\
			{\boldmath$n_{\rm run}    $} & $-0.0059\pm 0.0067$ & $-0.0108^{+0.0089}_{-0.0073}$ &   $-0.0079\pm 0.0070$ &  $-0.025\pm 0.013 $ \\
			{\boldmath$B      $}& --  & $8.6\times 10^{-5}{\; } ^{+1.8\times 10^{-5}}_{-9.3\times 10^{-5}}$ & $2.75\times 10^{-8}{\; }^{+5.5\times 10^{-9}}_{-2.8\times 10^{-8}}$   &  $< 0.766$  \\
			\hline
		\end{tabular}
	}
	\caption{Estimation for $n_{\rm run}$, $n_s$ and ${\rm{ln}}(10^{10} A_s) $ in $\Lambda$CDM and the CSLIM using the three schemes of Eq. \eqref{eq:params12y3}, along with the $B$ (in the appropriate units) parameter for each scheme. Mean value estimations are reported together with
		an error corresponding to 68\% confidence levels.}
	\label{tab:Bcero-planck_running}
\end{table*}

From Fig. \ref{fig:planck_running-vs-paramsconrun} and Table \ref{tab:Bcero-planck_running}, we can partially conclude that the effect of adding the $B$ parameter of the CSLIM is to increase $|n_{\rm run}|$ in all three cases. Interestingly, scheme II, i.e. $\lambda_k = \lambda_0 (k + B/k)$ yields a very similar value of $n_{\rm run}$ as the reference model. This means that, at least for this scheme of the CSLIM, the HFF could take the values $\epsilon_{1 \diamond} \simeq \epsilon_{2 \diamond} \simeq 10^{-2} > \epsilon_{3 \diamond}$, which is consistent with the hierarchy of the HFF. On the other hand, as we have mentioned in the Introduction, these same values for the HFF could not be achieved in the standard $\Lambda$CDM model because of the relations between $n_s$, $n_{\rm run}$ and the tensor-to-scalar ratio $r$. 

Finally, it is important to mention that the analysis we have presented included one additional parameter than the $\Lambda$CDM model with running, that is, the $B$ parameter of the CSLIM. In the next section, we will consider a slightly different version of the analysis in which the number of inflationary parameters will be the same in both models.

%

\section{Singling out the effects of the CSL inflationary model}\label{Sec4}

In the introduction we have mentioned that a confirmed detection of the running of the spectral index of order $|n_{\rm run}| \simeq 10^{-3}$, and lower bounds on the tensor-to-scalar ratio, e.g. $r \lesssim 10^{-3}$ may jeopardize slow roll inflation due to a possible violation of the hierarchy of the HFF. However, we argued that the CSLIM may introduce a similar effect as the running, in the primordial power spectrum, through $C(k)$ because of the $B$ parameter.

In order to test our hypothesis further, we will perform a comparison between the reference model (with and without running) and the CSLIM with $n_{\rm run} = 0$. The motivation for this analysis is twofold: first, we isolate the effect of $n_{\rm run}$ and $C(k)$ in each model, and second, we will have the same number of inflationary parameters in the reference model with running and in the CSLIM with no running.

\subsection{The effect of $C(k)$}

\begin{figure}
	\centering
	\includegraphics[width=\columnwidth]{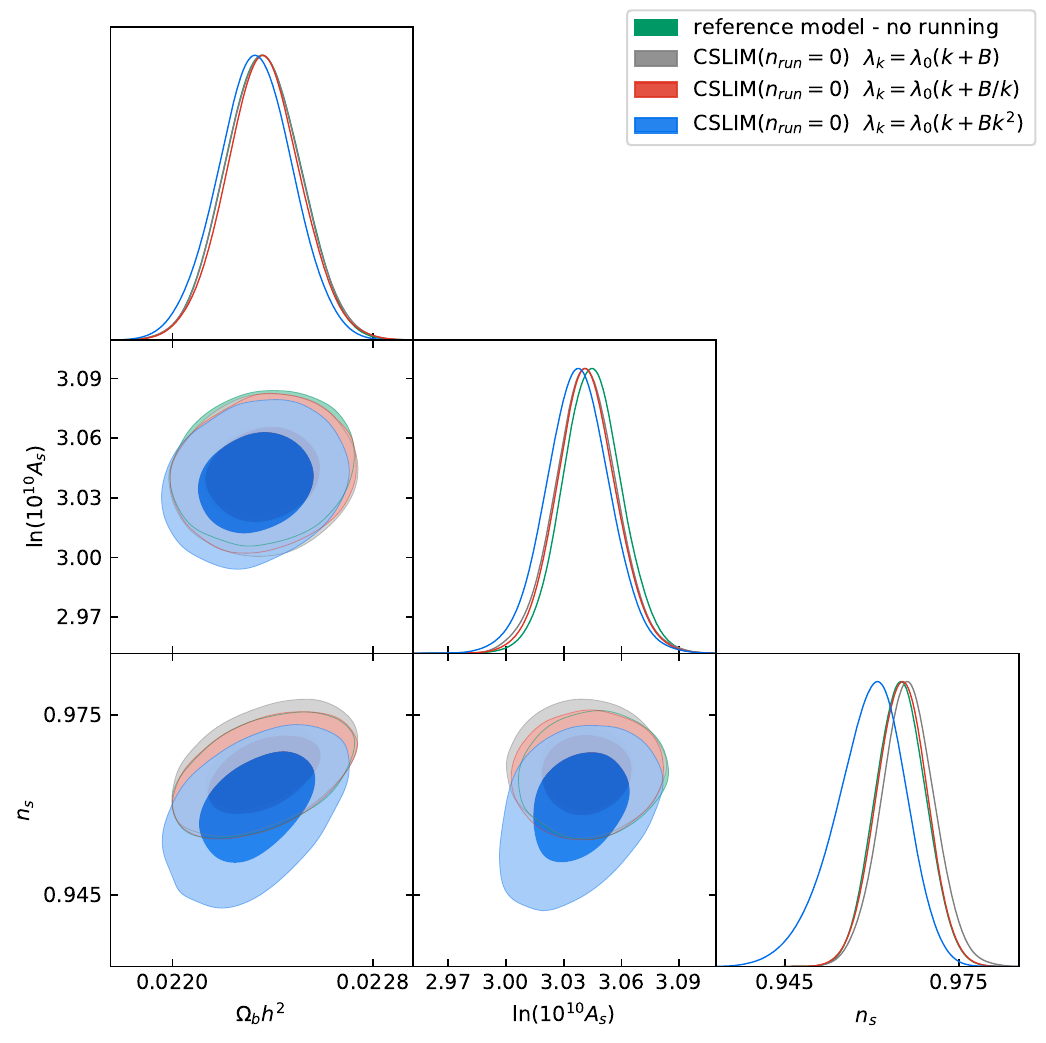}
	\caption{The effect of turning off the running of the scalar spectral index in both the reference and the CSLIM. The $68\%$ and $95\%$ confidence level contours are plotted together with the posterior probability. The three  schemes presented in \eqref{eq:params12y3} are used.}
	\label{fig:norunning}
\end{figure}

In Eq. \eqref{pdek}, we have obtained the modification of the standard prediction of $\mP_s (k)$ up to the second order in the HFF due to the incorporation of the CSL mechanism during inflation. Therefore, to analyze the effect of the CSLIM, without considering the running of the scalar spectral index, we compare a reference model with the three schemes in \eqref{eq:params12y3}, fixing $n_{\rm run} = 0$. In Fig. \ref{fig:norunning}, we display the parameters affected the most as a result of the CSLIM. This demonstrates that the CSLIM introduces unique observational features, and the magnitude of the effect, becomes influenced by the functional dependence of $\lambda_k$. 

\begin{figure}
	\includegraphics[width=\columnwidth]{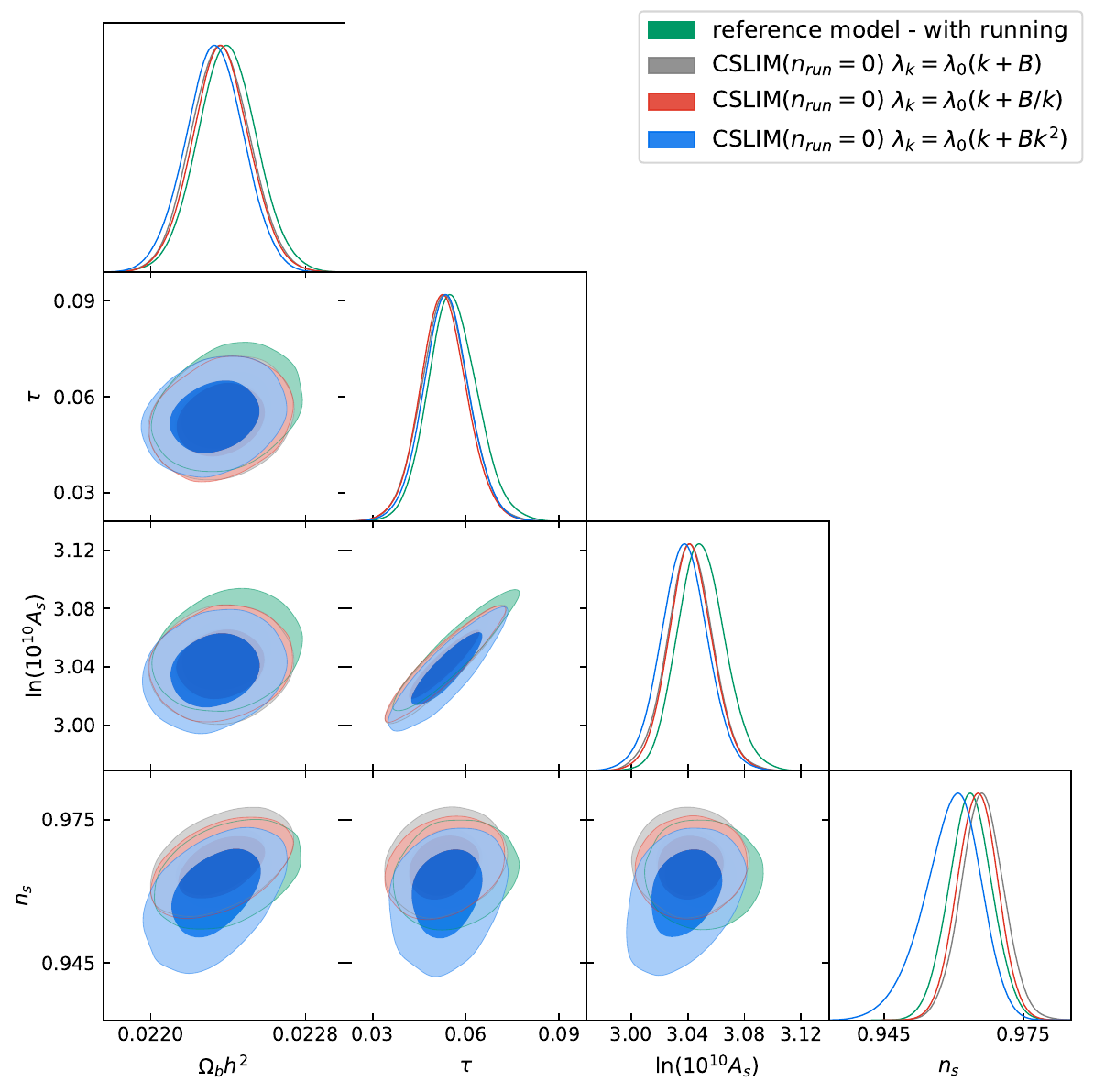}
	\caption{Comparison between the three different schemes characterizing  $\lambda_k$, all with $n_{\rm run} = 0$, and the reference $\Lambda$CDM model {\em with non-vanishing} running. The plot displays the {\em posteriori} probability with $68\%$ and $95\%$ confidence level contours for $\Omega_b h^2$, $\tau$, $\ln(10^{10}A_s)$, and $n_s$, which exhibit slight but compatible deviations from the reference model. Other parameters that complement the cosmological model show no difference at all. }
	
	\label{fig:planckrun-cslimnorun}
\end{figure}

Next, we aim to explore whether the CSLIM without running, i.e., $n_{\rm run} = 0$, can reproduce the parameter estimations of the $\Lambda$CDM + {\em running} model. Figure \ref{fig:planckrun-cslimnorun} illustrates that schemes I and II are perfectly compatible with observational data. However,  scheme III exhibits a slightly different behavior in the estimation of $n_s$, but we think that this variation does not fundamentally invalidate our initial premise.
{These results seem to indicate that the CSLIM can successfully fit the CMB data with $n_{\rm run}=0$, and simultaneously preserve the hierarchy  of the HFF. Additionally, in this case, we have employed the same number of parameters as in the standard $\Lambda$CDM inflationary model.
	This latter described by the usual set of parameters ($\Omega_b h^2$, $\Omega_c h^2$, $100\theta_{MC}$, $\tau$, ${\rm{ln}}(10^{10} A_s)$ and $n_s$)
	plus the running of the scalar spectral index, while our CSLIM
	model fixes  $n_{\rm run} = 0$ and includes $B$, which is direcly
	associated to the collapse parameter $\lambda_k$.
}

The estimations for the parameter $B$ in each case are shown in Table \ref{tab:Bnorunning}. The rest of the cosmological parameters show a very slight difference with no statistical significance (which is the case of $\tau$ and $\Omega_b h^2$) or no difference at all.

{In order to intuitively understand the results in Table 3, we introduce the following definitions: For scheme I, we define $\beta_1 \equiv B/k_\diamond$; for scheme II, $\beta_2 \equiv B/k_\diamond^2$; and for scheme III, $\beta_3 \equiv B k_\diamond$. 
	We note that, unlike the $B$ parameter, which has different dimensions in each scheme, the  $\beta_i$'s (where $i=1,2,3$) are completely dimensionless. Additionally, we have introduced the pivot scale $k_\diamond$ in the definitions as a reference scale.  We have shown in Sec. \ref{Sec:ref_model}  that the  form $\lambda_k = \lambda_0 k$ possesses no distinguishable observational features from the standard $\Lambda$CDM model. 
	Therefore, the parameters $\beta_i$ quantitatively represent the departure from the simplest functional dependence. Using the values from Table \ref{tab:Bnorunning}, we find that $\beta_1 \simeq 7.3 \times 10^{-4}$, $\beta_2 \simeq 8 \times 10^{-6}$, and $\beta_3 < 6.5 \times 10^{-3}$. Consequently, scheme III allows for a larger departure from $\lambda_k = \lambda_0 k$ compared to the other two schemes, while scheme II is the most restrictive.}

\begin{table}
	\centering
	{\setstretch {1.6}
		\begin{tabular}{lc}
			{\bf Schemes} & {\boldmath $B$}\\
			\hline
			{$\lambda_k=\lambda_0 (k+B)$} & $3.67 \times 10^{-5}{\; }^{+9.2\times 10^{-6}}_{-4.4\times 10^{-5}}$\\
			{$\lambda_k=\lambda_0 (k+B/k) $} & $2.02\times 10^{-8}{\; }^{+3.5\times 10^{-9}}_{-2.0\times 10^{-8}}$\\
			{$ \lambda_k=\lambda_0 (k+Bk^2) $} & $< 0.131                   $\\
			\hline
		\end{tabular}
	}
	\caption{Mean and $68\%$ limits for the estimation of $B$ employing the three 
		schemes characterizing $\lambda_k$ in \eqref{eq:params12y3}. The free parameter of the CSLIM is estimated while considering no running of the spectral index, along with the usual set of parameters in the $\Lambda$CDM cosmological model.}
		\label{tab:Bnorunning}
\end{table}





\section{Conclusions}\label{Sec5}

In this work, we have analyzed the previous theoretical calculations obtained from applying the {\em Continuous Spontaneous Localization} (CSL) model (or simply collapse mechanism) to the inflationary stage of the early Universe. For this purpose we have used the latest available CMB observational data. The primary aim of the CSL inflationary model is to explain the emergence of the cosmic seeds of large-scale structures. This is achieved purely by the dynamics of the collapse mechanism, which modifies  the initial state of the inflaton, namely, the perfectly homogeneous and isotropic Bunch-Davies vacuum state. The breakdown of the initial symmetries of the vacuum state occurs due to the evolution dictated by the modified Schr\"odigner's equation that is a central part of the CSL model. 
Thereby, addressing the measurement problem of quantum mechanics in the cosmological context.

The primary motivation of this article was to investigate whether the predictions of the CSL inflationary model, specifically concerning the primordial power spectrum at second order in the Hubble Flow Functions (HFF), can offer potential solutions to emerging observational challenges that the standard cosmological model might not address adequately. In particular, when considering the traditional slow roll inflationary model. 

From the latest data of the Cosmic Microwave Background (CMB) provided by the {\em Planck} collaboration, recent estimations suggest that the first logarithmic derivative of the scalar spectral index $n_s$ (with respect to $k$), or ``running'' of the spectral index $n_{\rm run}$, is consistent with a value of zero. However, the estimated value of the running is not precisely centered around zero, indicating that non-zero values might be favored in the future as more precise measurements and experiments become available.

At first glance, this may seem to imply that an additional parameter, $n_{\rm run}$, needs to be added to the $\Lambda$CDM model. However, if the absence of a confirmed detection of primordial gravitational waves persists and the bounds on the tensor-to-scalar ratio $r$ are lowered, then it could also indicate that the hierarchy of the HFF would be disrupted, which could pose a potential challenge for the standard slow roll inflationary paradigm \cite{runninglewis}.  {This is mainly because in traditional slow-roll inflation, the spectral index $n_s = 1-2\epsilon_{1 \diamond}-\epsilon_{2 \diamond}$, the running of the spectral index $n_{\rm run} = -2 \epsilon_{1 \diamond} \epsilon_{2 \diamond} - \epsilon_{2 \diamond} \epsilon_{3 \diamond}$, and the tensor-to-scalar ratio $r = 16 \epsilon_{1 \diamond}$ are related in a very specific manner: $n_{\rm run} = (n_s-1 + r/8)(r/8 + \epsilon_{3 \diamond})$. Currently, measurements of the spectral index are narrowly constrained \citep{Planck18b}, yielding $|n_s-1| \simeq 10^{-2}$. Therefore, if $r$ remains low enough to be undetected and future measurements of the running spectral index establish $|n_{\rm run}| \simeq 10^{-3}$ \citep{Bahr2021, Bahr2022}, it would constrain $\epsilon_{3 \diamond} \simeq 10^{-1}$. This, in turn, implies that $|\epsilon_{1 \diamond}| < |\epsilon_{2 \diamond}| < |\epsilon_{3 \diamond}|$, hence disrupting the hieararchy of the HFF. }

{On the other hand, in the CSL inflationary model, the predicted value of $r$ is generically small\footnote{This prediction is derived from the CSL collapse mechanism combined with the semiclassical gravity framework. Physically, the model requires  a source for the so-called tensor modes; this source is at the second order in the scalar perturbations of the inflaton. This contrasts with the standard paradigm in which the tensor modes are directly quantized, and no source is needed for their generation\citep{Lucila15, Maj17}. }, approximately $r \simeq 10^{-7} \epsilon_{1 \diamond}^2$ \citep{Lucila15, Maj17}.  This means that a value of $\epsilon_{1 \diamond} \simeq 10^{-2}$ is compatible with a very small $r$ (in contrast with the traditional picture). The predictions for the other two inflationary parameters, $n_s$ and $n_{\rm run}$, remain the same as in standard slow-roll inflation but are decoupled from $r$, unlike the standard prediction.  Additionally, in our model, current observational bounds on $n_s$ are also consistent with $\epsilon_{2 \diamond} \lesssim \epsilon_{1 \diamond}$. Another important feature of our model is that the familiar form of the primordial power spectrum is modified by an additional factor $C(k)$, which encodes information about the collapse mechanism through the parameter $\lambda_k$, as seen in Eq. \eqref{defCK2}. The main goal of this article was to analyze whether all these elements of the CSL inflationary model can be useful in addressing the potential challenges described in the previous paragraph. Specifically, we investigated whether the effect of the ``running'' of the spectral index could be mimicked by the function $C(k)$ while maintaining the hierarchy of the HFF, namely $|\epsilon_{1 \diamond}| > |\epsilon_{2 \diamond}| > |\epsilon_{3 \diamond}|$.}

%
%

To achieve our objective, we conducted a comprehensive analysis by comparing the theoretical predictions of our model with observational data. To do this, we performed a series of Markov--Monte Carlo chains for cosmological parameter estimation, adapting the {\sc cosmomc} and {\sc camb} source codes accordingly. Moreover, we considered three different functional forms of $\lambda_k$, which we called ``schemes'', see Eq. \eqref{eq:params12y3}. We have also considered a reference model based on the standard $\Lambda$CDM cosmological model. The results of this comparison between the reference and our modified model, which includes the $n_{\rm run}$ parameter, { demonstrate a high degree of compatibility with current observational data}, see Fig. \ref{fig:planck_running-vs-paramsconrun}.

In the last part of our analysis, we conducted a comparison between the reference $\Lambda$CDM cosmological model, which includes the $n_{\rm run}$ parameter, and our CSL inflationary model with $n_{\rm run} = 0$. By setting $n_{\rm run} = 0$ in our model, we isolated the effects of the collapse mechanism that could mimic a ``running'', while also ensuring that both models have an equal number of inflationary parameters. Specifically, the reference model includes a running index $n_{\rm run}$ in addition to the spectrum's amplitude $A_s$ and spectral index $n_s$ to characterize the primordial power spectra. In contrast, our CSL inflationary model --with no running-- achieves an equally good fit to observations using the same number of parameters, but in this case, we have $A_s$, $n_s$, and $B$, where the latter is directly associated with the collapse parameter $\lambda_k$. We performed this analysis for the three schemes characterizing $\lambda_k$ shown in Eq. \eqref{eq:params12y3}. The results obtained suggest that the CSL inflationary model can successfully accommodate the CMB data with $n_{\rm run}=0$, while maintaining the hierarchy of HFF, see Fig \ref{fig:planckrun-cslimnorun}. 
Future measurements of the CMB with increased precision will be crucial in strengthening the validity of our results. Furthermore, our study demonstrates that addressing the quantum measurement problem in cosmology leads to direct observational consequences.


\acknowledgments

M.P.P. and G.L. are supported by CONICET (Argentina). Also, M.P.P. and G.L. acknowledge support from the following project grants: Universidad Nacional de La Plata I+D  G175 and  PIP 11220200100729CO  CONICET (Argentina). M.P.P thanks Facultad de Ciencias Astron\'omicas y Geof\'{\i}sicas UNLP, as this work was mainly done under \textit{Programa de Retenci\'on de Recursos Humanos}.
{We are specially grateful to the anonymous referees for a helpful review. Their comments and suggestions have led to significant improvements in the presentation of the material in this manuscript.}



\bibliography{bibliografia}

\providecommand{\href}[2]{#2}\begingroup\raggedright\begin{thebibliography}{10}

\bibitem{Friedmann}
A.~{Friedmann}, {\it {{\"U}ber die Kr{\"u}mmung des Raumes}},  {\em Zeitschrift
  fur Physik} {\bf 10} (Jan., 1922) 377--386.

\bibitem{Einstein}
A.~Einstein and W.~de~Sitter, {\it On the relation between the expansion and
  the mean density of the universe},  {\em Proceedings of the National Academy
  of Sciences} {\bf 18} (1932), no.~3 213--214,
  [\href{http://arxiv.org/abs/https://www.pnas.org/doi/pdf/10.1073/pnas.18.3.213}{{\tt
  https://www.pnas.org/doi/pdf/10.1073/pnas.18.3.213}}].

\bibitem{deSitter}
W.~{de Sitter}, {\it {Some further computations regarding nonstatic
  universes}},  {\em \bain} {\bf 6} (Aug., 1931) 141.

\bibitem{Hubble}
E.~{Hubble}, {\it {A Relation between Distance and Radial Velocity among
  Extra-Galactic Nebulae}},  {\em Proceedings of the National Academy of
  Science} {\bf 15} (Mar., 1929) 168--173.

\bibitem{Lemaitre}
G.~{Lema{\^{\i}}tre}, {\it {A homogeneous universe of constant mass and
  increasing radius accounting for the radial velocity of extra-galactic
  nebulae}},  {\em \mnras} {\bf 91} (Mar., 1931) 483--490.

\bibitem{Slipher}
V.~M. {Slipher}, {\it {Nebulae}},  {\em Proceedings of the American
  Philosophical Society} {\bf 56} (Jan., 1917) 403--409.

\bibitem{Henrietta}
H.~S. {Leavitt} and E.~C. {Pickering}, {\it {Periods of 25 Variable Stars in
  the Small Magellanic Cloud.}},  {\em Harvard College Observatory Circular}
  {\bf 173} (Mar., 1912) 1--3.

\bibitem{Feeney2017}
S.~M. Feeney, D.~J. Mortlock, and N.~Dalmasso, {\it {Clarifying the Hubble
  constant tension with a Bayesian hierarchical model of the local distance
  ladder}},  {\em Mon. Not. Roy. Astron. Soc.} {\bf 476} (2018), no.~3
  3861--3882, [\href{http://arxiv.org/abs/1707.0000}{{\tt arXiv:1707.0000}}].

\bibitem{Planck18c}
{\bf Planck} Collaboration, N.~Aghanim et~al., {\it {Planck 2018 results. VI.
  Cosmological parameters}},  {\em Astron. Astrophys.} {\bf 641} (2020) A6,
  [\href{http://arxiv.org/abs/1807.0620}{{\tt arXiv:1807.0620}}].

\bibitem{eBOSS}
{\bf eBOSS} Collaboration, S.~Alam et~al., {\it {Completed SDSS-IV extended
  Baryon Oscillation Spectroscopic Survey: Cosmological implications from two
  decades of spectroscopic surveys at the Apache Point Observatory}},  {\em
  Phys. Rev. D} {\bf 103} (2021), no.~8 083533,
  [\href{http://arxiv.org/abs/2007.0899}{{\tt arXiv:2007.0899}}].

\bibitem{Guth81}
A.~H. Guth, {\it {The Inflationary Universe: A Possible Solution to the Horizon
  and Flatness Problems}},  {\em Phys. Rev.} {\bf D23} (1981) 347--356.

\bibitem{Hawking82}
S.~W. Hawking, {\it {The Development of Irregularities in a Single Bubble
  Inflationary Universe}},  {\em Phys. Lett.} {\bf 115B} (1982) 295.

\bibitem{Mukhanov}
V.~Mukhanov, {\em {Physical Foundations of Cosmology}}.
\newblock Cambridge Univ. Press, Cambridge, 2005.

\bibitem{penziaswilson}
A.~A. {Penzias} and R.~W. {Wilson}, {\it {A Measurement of Excess Antenna
  Temperature at 4080 Mc/s.}},  {\em \apj} {\bf 142} (July, 1965) 419--421.

\bibitem{cobe}
C.~Bennett, N.~Boggess, E.~Cheng, M.~Hauser, T.~Kelsall, J.~Mather, S.~Moseley,
  T.~Murdock, R.~Shafer, R.~Silverberg, G.~Smooth, R.~Weiss, and E.~Wright,
  {\it Scientific results from cobe},  {\em Advances in Space Research} {\bf
  13} (1993), no.~12 409--423.

\bibitem{wmap9cls}
C.~L. {Bennett}, D.~{Larson}, J.~L. {Weiland}, N.~{Jarosik}, G.~{Hinshaw},
  N.~{Odegard}, K.~M. {Smith}, R.~S. {Hill}, B.~{Gold}, M.~{Halpern},
  E.~{Komatsu}, M.~R. {Nolta}, L.~{Page}, D.~N. {Spergel}, E.~{Wollack},
  J.~{Dunkley}, A.~{Kogut}, M.~{Limon}, S.~S. {Meyer}, G.~S. {Tucker}, and
  E.~L. {Wright}, {\it {Nine-year Wilkinson Microwave Anisotropy Probe (WMAP)
  Observations: Final Maps and Results}},  {\em \apjs} {\bf 208} (Oct., 2013)
  20, [\href{http://arxiv.org/abs/1212.5225}{{\tt arXiv:1212.5225}}].

\bibitem{Planck18a}
{\bf Planck} Collaboration, N.~Aghanim et~al., {\it {Planck 2018 results. I.
  Overview and the cosmological legacy of Planck}},  {\em Astron. Astrophys.}
  {\bf 641} (2020) A1, [\href{http://arxiv.org/abs/1807.0620}{{\tt
  arXiv:1807.0620}}].

\bibitem{Planck18b}
{\bf Planck} Collaboration, Y.~Akrami et~al., {\it {Planck 2018 results. X.
  Constraints on inflation}},  {\em Astron. Astrophys.} {\bf 641} (2020) A10,
  [\href{http://arxiv.org/abs/1807.0621}{{\tt arXiv:1807.0621}}].

\bibitem{PSS06}
A.~Perez, H.~Sahlmann, and D.~Sudarsky, {\it {On the quantum origin of the
  seeds of cosmic structure}},  {\em Class. Quant. Grav.} {\bf 23} (2006)
  2317--2354, [\href{http://arxiv.org/abs/gr-qc/0508100}{{\tt gr-qc/0508100}}].

\bibitem{Shortcomings}
D.~Sudarsky, {\it {Shortcomings in the Understanding of Why Cosmological
  Perturbations Look Classical}},  {\em Int.\ J.\ Mod.\ Phys.\ D} {\bf 20}
  (2011) 509--552, [\href{http://arxiv.org/abs/0906.0315}{{\tt
  arXiv:0906.0315}}].

\bibitem{Susana13}
S.~Landau, G.~Le\'on, and D.~Sudarsky, {\it {Quantum Origin of the Primordial
  Fluctuation Spectrum and its Statistics}},  {\em Phys. Rev. D} {\bf 88}
  (2013), no.~2 023526, [\href{http://arxiv.org/abs/1107.3054}{{\tt
  arXiv:1107.3054}}].

\bibitem{Okon16}
E.~Okon and D.~Sudarsky, {\it {Less Decoherence and More Coherence in Quantum
  Gravity, Inflationary Cosmology and Elsewhere}},  {\em Found. Phys.} {\bf 46}
  (2016), no.~7 852--879, [\href{http://arxiv.org/abs/1512.0529}{{\tt
  arXiv:1512.0529}}].

\bibitem{Elias2021}
J.~Berjon, E.~Okon, and D.~Sudarsky, {\it {Critical review of prevailing
  explanations for the emergence of classicality in cosmology}},  {\em Phys.
  Rev. D} {\bf 103} (2021), no.~4 043521,
  [\href{http://arxiv.org/abs/2009.0999}{{\tt arXiv:2009.0999}}].

\bibitem{kiefer2}
C.~Kiefer, {\it {Origin of classical structure from inflation}},  {\em Nucl.
  Phys. Proc. Suppl.} {\bf 88} (2000) 255--258,
  [\href{http://arxiv.org/abs/astro-ph/0006252}{{\tt astro-ph/0006252}}].

\bibitem{polarski}
D.~Polarski and A.~A. Starobinsky, {\it {Semiclassicality and decoherence of
  cosmological perturbations}},  {\em Class. Quant. Grav.} {\bf 13} (1996)
  377--392, [\href{http://arxiv.org/abs/gr-qc/9504030}{{\tt gr-qc/9504030}}].

\bibitem{jmartin}
J.~{Martin}, V.~{Vennin}, and P.~{Peter}, {\it {Cosmological inflation and the
  quantum measurement problem}},  {\em Phys.Rev. D} {\bf 86} (Nov., 2012)
  103524, [\href{http://arxiv.org/abs/1207.2086}{{\tt arXiv:1207.2086}}].

\bibitem{jmartinPRL}
J.~Martin and V.~Vennin, {\it {A cosmic shadow on CSL}},  {\em Phys. Rev.
  Lett.} {\bf 124} (2020), no.~8 080402,
  [\href{http://arxiv.org/abs/1906.0440}{{\tt arXiv:1906.0440}}].

\bibitem{pedrocsl}
P.~Ca\~{n}ate, P.~Pearle, and D.~Sudarsky, {\it {Continuous spontaneous
  localization wave function collapse model as a mechanism for the emergence of
  cosmological asymmetries in inflation}},  {\em Phys. Rev.} {\bf D87} (2013),
  no.~10 104024, [\href{http://arxiv.org/abs/1211.3463}{{\tt
  arXiv:1211.3463}}].

\bibitem{Das2013}
S.~Das, K.~Lochan, S.~Sahu, and T.~P. Singh, {\it {Quantum to classical
  transition of inflationary perturbations: Continuous spontaneous localization
  as a possible mechanism}},  {\em Phys. Rev.} {\bf D88} (2013), no.~8 085020,
  [\href{http://arxiv.org/abs/1304.5094}{{\tt arXiv:1304.5094}}]. [Erratum:
  Phys. Rev.D89,no.10,109902(2014)].

\bibitem{kiefer}
C.~Kiefer and D.~Polarski, {\it {Why do cosmological perturbations look
  classical to us?}},  {\em Adv. Sci. Lett.} {\bf 2} (2009) 164--173,
  [\href{http://arxiv.org/abs/0810.0087}{{\tt arXiv:0810.0087}}].

\bibitem{pintoneto}
N.~Pinto-Neto, G.~Santos, and W.~Struyve, {\it {Quantum-to-classical transition
  of primordial cosmological perturbations in de Broglie--Bohm quantum
  theory}},  {\em Phys. Rev.} {\bf D85} (2012) 083506,
  [\href{http://arxiv.org/abs/1110.1339}{{\tt arXiv:1110.1339}}].

\bibitem{valentini}
A.~Valentini, {\it {Inflationary Cosmology as a Probe of Primordial Quantum
  Mechanics}},  {\em Phys. Rev.} {\bf D82} (2010) 063513,
  [\href{http://arxiv.org/abs/0805.0163}{{\tt arXiv:0805.0163}}].

\bibitem{goldstein}
S.~Goldstein, W.~Struyve, and R.~Tumulka, {\it {The Bohmian Approach to the
  Problems of Cosmological Quantum Fluctuations}},  2015.

\bibitem{Stephon16}
S.~Alexander, D.~Jyoti, and J.~Magueijo, {\it {Inflation and the quantum
  measurement problem}},  {\em Phys. Rev. D} {\bf 94} (2016), no.~4 043502,
  [\href{http://arxiv.org/abs/1602.0121}{{\tt arXiv:1602.0121}}].

\bibitem{Picci19}
M.~P. {Piccirilli}, G.~{Le{\'o}n}, S.~J. {Landau}, M.~{Benetti}, and
  D.~{Sudarsky}, {\it {Constraining quantum collapse inflationary models with
  current data: The semiclassical approach}},  {\em International Journal of
  Modern Physics D} {\bf 28} (Jan, 2019) 1950041,
  [\href{http://arxiv.org/abs/1709.0623}{{\tt arXiv:1709.0623}}].

\bibitem{Leon2020}
G.~R. Bengochea, G.~Le\'on, P.~Pearle, and D.~Sudarsky, {\it {Discussions about
  the landscape of possibilities for treatments of cosmic inflation involving
  continuous spontaneous localization models}},  {\em Eur. Phys. J. C} {\bf 80}
  (2020), no.~11 1021, [\href{http://arxiv.org/abs/2008.0528}{{\tt
  arXiv:2008.0528}}].

\bibitem{Leon2021}
G.~Le\'on and G.~R. Bengochea, {\it {Enlightening the CSL model landscape in
  inflation}},  {\em Eur. Phys. J. C} {\bf 81} (2021), no.~12 1055,
  [\href{http://arxiv.org/abs/2107.0547}{{\tt arXiv:2107.0547}}].

\bibitem{ashtekar2020}
A.~Ashtekar, A.~Corichi, and A.~Kesavan, {\it {Emergence of classical behavior
  in the early universe}},  {\em Phys. Rev. D} {\bf 102} (2020), no.~2 023512,
  [\href{http://arxiv.org/abs/2004.1068}{{\tt arXiv:2004.1068}}].

\bibitem{Bell81}
J.~S. Bell, {\em Quantum Mechanics for cosmologists}.
\newblock Quantum Gravity 2. eds. Isham, C., Penrose, R. and Sciama, D., Oxford
  University Press, 1981.

\bibitem{Maudlin95}
T.~Maudlin, {\it Three measurement problems},  {\em Topoi} {\bf 14} (1995)
  7--15.

\bibitem{Ghirardi86}
G.~Ghirardi, A.~Rimini, and T.~Weber, {\it {A Unified Dynamics for Micro and
  MACRO Systems}},  {\em Phys.Rev.} {\bf D34} (1986) 470.

\bibitem{Pearle89}
P.~M. Pearle, {\it {Combining Stochastic Dynamical State Vector Reduction With
  Spontaneous Localization}},  {\em Phys.Rev.} {\bf A39} (1989) 2277--2289.

\bibitem{Diosi84}
L.~{Di{\'o}si}, {\it {Gravitation and quantum-mechanical localization of
  macro-objects}},  {\em Physics Letters A} {\bf 105} (Oct., 1984) 199--202,
  [\href{http://arxiv.org/abs/1412.0201}{{\tt arXiv:1412.0201}}].

\bibitem{Penrose96}
R.~Penrose, {\it {On gravity's role in quantum state reduction}},  {\em
  Gen.Rel.Grav.} {\bf 28} (1996) 581--600.

\bibitem{Bassi1}
A.~{Bassi} and G.~{Ghirardi}, {\it {Dynamical reduction models}},  {\em Phys.
  Rept.} {\bf 379} (Jun, 2003) 257--426,
  [\href{http://arxiv.org/abs/quant-ph/0302164}{{\tt quant-ph/0302164}}].

\bibitem{Bassi2}
A.~{Bassi}, K.~{Lochan}, S.~{Satin}, T.~P. {Singh}, and H.~{Ulbricht}, {\it
  {Models of wave-function collapse, underlying theories, and experimental
  tests}},  {\em Reviews of Modern Physics} {\bf 85} (Apr, 2013) 471--527,
  [\href{http://arxiv.org/abs/1204.4325}{{\tt arXiv:1204.4325}}].

\bibitem{Pearle1995}
P.~M. Pearle and E.~Squires, {\it {Gravity, energy conservation and parameter
  values in collapse models}},  {\em Found. Phys.} {\bf 26} (1996) 291,
  [\href{http://arxiv.org/abs/quant-ph/9503019}{{\tt quant-ph/9503019}}].

\bibitem{PearleMisc}
P.~{Pearle}, {\it {Collapse Miscellany}},
  \href{http://arxiv.org/abs/1209.5082}{{\tt arXiv:1209.5082}}.

\bibitem{Pearle06A}
P.~{Pearle}, {\it {How stands collapse I}},  {\em Journal of Physics A
  Mathematical General} {\bf 40} (Mar., 2007) 3189--3204,
  [\href{http://arxiv.org/abs/quant-ph/0611211}{{\tt quant-ph/0611211}}].

\bibitem{Bassiexp21}
G.~Gasbarri, A.~Belenchia, M.~Carlesso, S.~Donadi, A.~Bassi, R.~Kaltenbaek,
  M.~Paternostro, and H.~Ulbricht, {\it {Testing the foundations of quantum
  physics in space Interferometric and non-interferometric tests with Large
  Particles}},  {\em Commun. Phys.} {\bf 4} (2021) 155,
  [\href{http://arxiv.org/abs/2106.0534}{{\tt arXiv:2106.0534}}].

\bibitem{Bassi2022}
M.~Carlesso, S.~Donadi, L.~Ferialdi, M.~Paternostro, H.~Ulbricht, and A.~Bassi,
  {\it {Present status and future challenges of non-interferometric tests of
  collapse models}},  {\em Nature Phys.} {\bf 18} (2022), no.~3 243--250,
  [\href{http://arxiv.org/abs/2203.0423}{{\tt arXiv:2203.0423}}].

\bibitem{BassiCMB}
K.~Lochan, S.~Das, and A.~Bassi, {\it {Constraining CSL strength parameter
  $\lambda$ from standard cosmology and spectral distortions of CMBR}},  {\em
  Phys. Rev. D} {\bf 86} (2012) 065016,
  [\href{http://arxiv.org/abs/1206.4425}{{\tt arXiv:1206.4425}}].

\bibitem{Leon16}
G.~Leon and G.~R. Bengochea, {\it {Emergence of inflationary perturbations in
  the CSL model}},  {\em Eur. Phys. J.} {\bf C76} (2016), no.~1 29,
  [\href{http://arxiv.org/abs/1502.0490}{{\tt arXiv:1502.0490}}].

\bibitem{Maj17}
G.~Le\'on, A.~Majhi, E.~Okon, and D.~Sudarsky, {\it {Reassessing the link
  between B-modes and inflation}},  {\em Phys. Rev. D} {\bf 96} (2017), no.~10
  101301, [\href{http://arxiv.org/abs/1607.0352}{{\tt arXiv:1607.0352}}].

\bibitem{Beneficios}
E.~Okon and D.~Sudarsky, {\it {Benefits of Objective Collapse Models for
  Cosmology and Quantum Gravity}},  {\em Found. Phys.} {\bf 44} (2014)
  114--143, [\href{http://arxiv.org/abs/1309.1730}{{\tt arXiv:1309.1730}}].

\bibitem{Bassi2021}
A.~Gundhi, J.~L. Gaona-Reyes, M.~Carlesso, and A.~Bassi, {\it {Impact of
  Dynamical Collapse Models on Inflationary Cosmology}},  {\em Phys. Rev.
  Lett.} {\bf 127} (2021), no.~9 091302,
  [\href{http://arxiv.org/abs/2102.0768}{{\tt arXiv:2102.0768}}].

\bibitem{jmartinpotentials}
J.~Martin, C.~Ringeval, and V.~Vennin, {\it {Encyclop\ae{}dia Inflationaris}},
  {\em Phys. Dark Univ.} {\bf 5-6} (2014) 75--235,
  [\href{http://arxiv.org/abs/1303.3787}{{\tt arXiv:1303.3787}}].

\bibitem{jmartin2019}
D.~Chowdhury, J.~Martin, C.~Ringeval, and V.~Vennin, {\it {Assessing the
  scientific status of inflation after Planck}},  {\em Phys. Rev. D} {\bf 100}
  (2019), no.~8 083537, [\href{http://arxiv.org/abs/1902.0395}{{\tt
  arXiv:1902.0395}}].

\bibitem{Adshead2010}
P.~Adshead, R.~Easther, J.~Pritchard, and A.~Loeb, {\it {Inflation and the
  Scale Dependent Spectral Index: Prospects and Strategies}},  {\em JCAP} {\bf
  02} (2011) 021, [\href{http://arxiv.org/abs/1007.3748}{{\tt
  arXiv:1007.3748}}].

\bibitem{Bahr2021}
R.~Easther, B.~Bahr-Kalus, and D.~Parkinson, {\it {Running primordial
  perturbations: Inflationary dynamics and observational constraints}},  {\em
  Phys. Rev. D} {\bf 106} (2022), no.~6 L061301,
  [\href{http://arxiv.org/abs/2112.1092}{{\tt arXiv:2112.1092}}].

\bibitem{Planck15_inf}
{Planck Collaboration}, {\it {Planck 2015 results. XX. Constraints on
  inflation}},  {\em \aap} {\bf 594} (Sept., 2016) A20,
  [\href{http://arxiv.org/abs/1502.0211}{{\tt arXiv:1502.0211}}].

\bibitem{Bahr2022}
B.~Bahr-Kalus, D.~Parkinson, and R.~Easther, {\it {Constraining cosmic
  inflation with observations: Prospects for 2030}},  {\em Mon. Not. Roy.
  Astron. Soc.} {\bf 520} (2023), no.~2 2405--2416,
  [\href{http://arxiv.org/abs/2212.0411}{{\tt arXiv:2212.0411}}].

\bibitem{liddleSR}
A.~R. Liddle, P.~Parsons, and J.~D. Barrow, {\it {Formalizing the slow roll
  approximation in inflation}},  {\em Phys. Rev. D} {\bf 50} (1994) 7222--7232,
  [\href{http://arxiv.org/abs/astro-ph/9408015}{{\tt astro-ph/9408015}}].

\bibitem{runninglewis}
J.~P. Vieira, C.~T. Byrnes, and A.~Lewis, {\it {Can power spectrum observations
  rule out slow-roll inflation?}},  {\em JCAP} {\bf 01} (2018) 019,
  [\href{http://arxiv.org/abs/1710.0840}{{\tt arXiv:1710.0840}}].

\bibitem{running2020}
G.~Leon and M.~P. Piccirilli, {\it {Generation of inflationary perturbations in
  the continuous spontaneous localization model: The second order power
  spectrum}},  {\em Phys. Rev. D} {\bf 102} (2020), no.~4 043515,
  [\href{http://arxiv.org/abs/2006.0309}{{\tt arXiv:2006.0309}}].

\bibitem{Lucila15}
G.~Le\'on, L.~Kraiselburd, and S.~J. Landau, {\it {Primordial gravitational
  waves and the collapse of the wave function}},  {\em Phys. Rev. D} {\bf 92}
  (2015), no.~8 083516, [\href{http://arxiv.org/abs/1509.0839}{{\tt
  arXiv:1509.0839}}].

\bibitem{terreroHFF}
D.~J. Schwarz, C.~A. Terrero-Escalante, and A.~A. Garcia, {\it {Higher order
  corrections to primordial spectra from cosmological inflation}},  {\em Phys.
  Lett. B} {\bf 517} (2001) 243--249,
  [\href{http://arxiv.org/abs/astro-ph/0106020}{{\tt astro-ph/0106020}}].

\bibitem{terreroHFF2}
D.~J. Schwarz and C.~A. Terrero-Escalante, {\it {Primordial fluctuations and
  cosmological inflation after WMAP 1.0}},  {\em JCAP} {\bf 08} (2004) 003,
  [\href{http://arxiv.org/abs/hep-ph/0403129}{{\tt hep-ph/0403129}}].

\bibitem{alberto}
A.~Diez-Tejedor and D.~Sudarsky, {\it {Towards a formal description of the
  collapse approach to the inflationary origin of the seeds of cosmic
  structure}},  {\em JCAP} {\bf 1207} (2012) 045,
  [\href{http://arxiv.org/abs/1108.4928}{{\tt arXiv:1108.4928}}].

\bibitem{erandy}
P.~Ca\~{n}ate, E.~Ramirez, and D.~Sudarsky, {\it {Semiclassical Self Consistent
  Treatment of the Emergence of Seeds of Cosmic Structure. The second order
  construction}},  {\em JCAP} {\bf 1808} (2018) 043,
  [\href{http://arxiv.org/abs/1802.0223}{{\tt arXiv:1802.0223}}].

\bibitem{benito}
B.~A. Ju\'arez-Aubry, B.~S. Kay, and D.~Sudarsky, {\it {Generally covariant
  dynamical reduction models and the Hadamard condition}},  {\em Phys. Rev.}
  {\bf D97} (2018), no.~2 025010, [\href{http://arxiv.org/abs/1708.0937}{{\tt
  arXiv:1708.0937}}].

\bibitem{benito2019}
B.~A. Ju{\'a}rez-Aubry, T.~Miramontes, and D.~Sudarsky, {\it {Semiclassical
  theories as initial value problems}},  {\em J. Math. Phys.} {\bf 61} (2020),
  no.~3 032301, [\href{http://arxiv.org/abs/1907.0996}{{\tt arXiv:1907.0996}}].

\bibitem{Mukhanov81}
V.~F. Mukhanov and G.~V. Chibisov, {\it {Quantum Fluctuations and a Nonsingular
  Universe}},  {\em JETP Lett.} {\bf 33} (1981) 532--535. [Pisma Zh. Eksp.
  Teor. Fiz.33,549(1981)].

\bibitem{Daniel10}
G.~Leon and D.~Sudarsky, {\it {The Slow roll condition and the amplitude of the
  primordial spectrum of cosmic fluctuations: Contrasts and similarities of
  standard account and the 'collapse scheme'}},  {\em Class. Quant. Grav.} {\bf
  27} (2010) 225017, [\href{http://arxiv.org/abs/1003.5950}{{\tt
  arXiv:1003.5950}}].

\bibitem{Modak14}
S.~K. Modak, L.~Ort\'iz, I.~Pe\~na, and D.~Sudarsky, {\it {Black hole
  evaporation: information loss but no paradox}},  {\em General Relativity and
  Gravitation} {\bf 47} (Oct, 2015) 120,
  [\href{http://arxiv.org/abs/1406.4898}{{\tt arXiv:1406.4898}}].

\bibitem{Modak15}
S.~K. Modak, L.~Ort\'iz, I.~Pe\~na, and D.~Sudarsky, {\it {Nonparadoxical loss
  of information in black hole evaporation in a quantum collapse model}},  {\em
  Phys. Rev. D} {\bf 91} (Jun, 2015) 124009,
  [\href{http://arxiv.org/abs/1408.3062}{{\tt arXiv:1408.3062}}].

\bibitem{sandro2017}
K.~{Piscicchia}, A.~{Bassi}, C.~{Curceanu}, R.~{Grande}, S.~{Donadi},
  B.~{Hiesmayr}, and A.~{Pichler}, {\it {CSL Collapse Model Mapped with the
  Spontaneous Radiation}},  {\em Entropy} {\bf 19} (June, 2017) 319,
  [\href{http://arxiv.org/abs/1710.0197}{{\tt arXiv:1710.0197}}].

\bibitem{toros2016}
M.~Toro{\v s} and A.~Bassi, {\it {Bounds on quantum collapse models from
  matter-wave interferometry: calculational details}},  {\em J. Phys. A} {\bf
  51} (2018), no.~11 115302, [\href{http://arxiv.org/abs/1601.0293}{{\tt
  arXiv:1601.0293}}].

\bibitem{carlesso2016}
M.~Carlesso, A.~Bassi, P.~Falferi, and A.~Vinante, {\it {Experimental bounds on
  collapse models from gravitational wave detectors}},  {\em Phys. Rev. D} {\bf
  94} (2016), no.~12 124036, [\href{http://arxiv.org/abs/1606.0458}{{\tt
  arXiv:1606.0458}}].

\bibitem{cota_lambda0}
A.~Tilloy and T.~M. Stace, {\it {Neutron star heating constraints on
  wave-function collapse models}},  {\em Phys. Rev. Lett.} {\bf 123} (2019),
  no.~8 080402, [\href{http://arxiv.org/abs/1901.0547}{{\tt arXiv:1901.0547}}].

\bibitem{CAMB}
A.~Lewis, A.~Challinor, and A.~Lasenby, {\it {Efficient computation of CMB
  anisotropies in closed FRW models}},  {\em Astrophys. J.} {\bf 538} (2000)
  473--476, [\href{http://arxiv.org/abs/astro-ph/9911177}{{\tt
  astro-ph/9911177}}].

\bibitem{COSMOMC}
A.~{Lewis} and S.~{Bridle}, {\it {Cosmological parameters from CMB and other
  data: A Monte Carlo approach}},  {\em Physical Review D} {\bf 66} (Nov.,
  2002) 103511, [\href{http://arxiv.org/abs/astro-ph/0205436}{{\tt
  astro-ph/0205436}}].

\bibitem{Planck18like}
{Planck Collaboration}, {\it {Planck 2018 results. V. CMB power spectra and
  likelihoods}},  {\em \aap} {\bf 641} (Sept., 2020) A5,
  [\href{http://arxiv.org/abs/1907.1287}{{\tt arXiv:1907.1287}}].

\bibitem{Weinberg2008}
S.~Weinberg, {\em {Cosmology}}.
\newblock New York: Oxford University Press, 2008.

\bibitem{Goldberg91}
D.~Goldberg, {\it What every computer scientist should know about
  floating-point arithmetic.},  {\em ACM Comput. Surv.} {\bf 23} (1991), no.~1
  5--48. corrigendum: ACM Computing Surveys 23(3): 413 (1991), comments: ACM
  Computing Surveys 24(2): 319 (1992).

\end{thebibliography}\endgroup
\bibliographystyle{JHEP} 
%
%
%



\end{document}